\documentclass[twocolumn]{aastex701}
\newcommand{\simi}{$\sim$}
\newcommand{\rday}{~d$^{-1}$\xspace}

\newcommand{\sys}{SySts\xspace}
\newcommand{\axper}{AX~Per\xspace}
\newcommand{\cicyg}{CI~Cyg\xspace}
\newcommand{\zand}{Z~And\xspace}

\newcommand{\asas}{\text{ASAS-SN}\xspace}

\usepackage{graphicx}	
\usepackage{amsmath}	

\usepackage{xspace}
\usepackage{multirow}
\usepackage{url}
\usepackage{booktabs}
\usepackage{multirow}

\begin{document}

\title[Intermediate-Timescale Variabilities in SySts]{Photometric Modulations on Intermediate Timescales in the Symbiotic Binaries}
\footnote{Released on ..., ....}

\author[0000-0001-8697-2385]{Melis Yardımcı$^*$}
\affiliation{Department of Astronomy and Space Sciences, University of Ege, 35100, {\.I}zmir, Türkiye}
\email{melisyardimci@gmail.com}
\email{melis.yardimci@mail.ege.edu.tr}
\thanks{Corresponding author: melisyardimci@gmail.com; \\ melis.yardimci@mail.ege.edu.tr}

\author[0000-0002-8301-0604]{Samet Ok}
\affiliation{Department of Astronomy and Space Sciences, University of Ege, 35100, {\.I}zmir, Türkiye}
\email{samet.ok@gmail.com}

\author[0000-0001-6695-1157]{Belinda Kalomeni}
\affiliation{Department of Astronomy and Space Sciences, University of Ege, 35100, {\.I}zmir, Türkiye}
\email{}


\begin{abstract}
We present a multi-band study of three symbiotic binaries using combined ground- and space-based monitoring that spans up to 14 years. These datasets enable a systematic investigation of variability on intermediate timescales (tens of days) and the detection of shorter-period signals.
All systems display coherent photometric modulations that are distinct from the orbital cycles. In AX~Per, a dominant 75-day signal and its 37-day harmonic are identified, which we interpret as pulsations of the cool giant. CI~Cyg exhibits a stable modulation between 70 and 74~days, which likely arises from a combination of pulsation and circumstellar or disk-related variability. For Z~And, we confirm a persistent modulation between 55 and 60~days, consistent with semiregular pulsations of the cool component.
Additionally, space-based data reveal further short-period variability, including coherent signals at 26.7 and 66.6 minutes in Z~And and CI~Cyg, respectively, and a quasi-periodic modulation near 0.95~days in AX~Per. These detections suggest the presence of rapid activity driven by accretion or rotation, superposed on the intermediate timescale behavior. 
Our results show that the observed variability in these symbiotic binaries reflects the combined effects of cool giant pulsation, circumstellar or disk activity, and possible rotation of the hot component. The multi-timescale behavior revealed here offers new constraints on mass transfer and activity cycles in interacting binaries.
\end{abstract}




\section{Introduction} \label{sec:intro}
Symbiotic binaries (\sys) consist of an evolved red giant transferring mass to a compact hot companion, typically a white dwarf \citep{Belczynski00,Munari19,Merc25_rew}. 
Their light curves exhibit complex variability on multiple timescales: orbital modulations associated with eclipses and illumination effects \citep{Formiggini90}; long-term outbursts driven by changes in nuclear burning or accretion \citep{Mikolajewska02, Sokoloski06a,King09,Bollimpalli18}; and relatively short-term oscillations linked to pulsation, accretion instabilities \citep{Kenyon86,Mikolajewska95,Munari19}, or possible white dwarf rotation \citep{Sokoloski99,formiggini09}. 

Despite decades of study, the intermediate regime between pulsational and orbital variability ($\approx$30–200~d) remains poorly understood. Signals in this range have been reported sporadically in several systems \citep[e.g., \zand,][]{Sekeras19, Skopal18}, yet their physical origin and prevalence across symbiotic binaries have not been systematically investigated.

Photometric changes in symbiotic binaries arise from several interacting mechanisms that operate simultaneously. Orbital modulation produces eclipses or illumination effects as the red giant and ionized nebula move relative to the observer \citep{Formiggini90,Skopal05}. The cool giant often shows semi-regular pulsations that generate variability on timescales of tens to hundreds of days \citep{Kenyon86,Formiggini94,Mikolajewska03}. Mass-transfer fluctuations and instabilities in the accretion disk can introduce shorter-term variations, including flickering and quasi-periodic signals \citep{Sokoloski99,Luna07}. During active phases, enhanced accretion or expansion of the hot component reshapes the ionized nebula, producing complex multi-band light-curve behavior \citep{Skopal03,Sokoloski06a}. These combined processes make the interpretation of photometric variability challenging and highlight the need for long-term, multi-band monitoring to separate orbital, pulsational, and accretion-driven components.
The optical emission of symbiotic systems is a complex combination of radiation from three main components: the cool giant, the hot companion, and the surrounding ionized circumbinary nebula \citep{Muerset91, Skopal05}. Consequently, separating the relative contributions of these components using photometric data alone is inherently difficult, particularly during active phases.

\axper, \cicyg, and \zand each exhibit additional system-specific properties that make them valuable laboratories for studying multi-timescale variability. \axper is an eclipsing symbiotic binary with an orbital period of $682.1\pm1.4$~d \citep{Fekel00_II}, showing alternating quiescent and active phases in which its minima become distorted and asymmetric \citep{Formiggini94, Skopal01,Skopal11}. \cicyg, with an orbital period of $853.8\pm2.9$~d \citep{Fekel00_II}, hosts both a dense nebular region within the Roche lobe of the hot companion and a more extended bipolar structure \citep{Kenyon91}, and its photometric behavior is strongly influenced by these components and by variable mass transfer \citep{Mikolajewska03, Skopal05}. \zand, the prototype of symbiotic systems, has an orbital period of $759.0\pm1.9$~d \citep{Fekel00_II} and undergoes recurrent optical outbursts \citep{Skopal03}. 
Its hot component shows rapid brightness modulations at 
$1682.6 \pm 0.6$~s \citep[the quoted uncertainty reflects the formal statistical error;][]{Sokoloski99} 
interpreted as the magnetic white dwarf’s spin period, making it a key system for investigating accretion-driven variability.

Outbursts in symbiotic binaries are commonly grouped into several categories.
\zand-type outbursts exemplified by the recurrent activity of \zand itself, are characterized by a rapid rise in optical brightness followed by a slow decline that may persist for years or even decades \citep{Sokoloski06a, Bollimpalli18}.
Early interpretations attributed \zand-type outbursts either to accretion-disk instabilities similar to those in dwarf novae \citep{Kenyon86} or to the expansion and cooling of the hot component in response to enhanced accretion \citep{Mikolajewska95}. 
More recent studies suggest that \zand-type outbursts result from a transient increase in the accretion rate that exceeds the level required to sustain stable burning on the white dwarf surface \citep[e.g.,][]{Skopal17, Munari19, Skopal20}.
These results demonstrate that \zand-type outbursts can also occur in systems previously regarded as quiescent and underscore the importance of long-term photometric monitoring for constraining their physical origin.
Symbiotic or slow novae evolve more gradually and are powered by thermonuclear flashes on the white dwarf surface \citep{Mikolajewska02, Kato23}. Recurrent novae, such as RS~Oph, occur in systems with massive white dwarfs that undergo repeated thermonuclear explosions on timescales of years to decades \citep{Sokoloski06b, Luna07}. These diverse outburst mechanisms illustrate the range of accretion and nuclear processes influencing the optical variability of symbiotic binaries and highlight the need to investigate how such activity couples to variability on intermediate and short timescales.

In this study, we perform a unified analysis of 14~yr of multiband ground-based photometry for \axper, \cicyg, and \zand, combined with complementary high-cadence observations from TESS. We aim to identify and compare intermediate-timescale modulations across the three systems, to assess the stability and physical origin of these signals, and to search for short-period oscillations that may trace accretion dynamics or white dwarf rotation. 
Taken together, these datasets allow us to investigate variability over a wide range of timescales in these prototypical nuclear-powered systems. Our results suggest that such multi-timescale behavior is a prominent feature of this class of symbiotic binaries, reflecting the complex interactions inherent to their active phases.

\begin{table}
\centering
\Huge
\caption{Observation log of \axper, \cicyg and \zand.}
\resizebox{\columnwidth}{!}{
\begin{tabular}{llcccc}
\hline
\hline
\textbf{Name} & \textbf{Observatory} & \textbf{Filter/Sector} & \textbf{Starting} & \textbf{Ending} \\ 
\hline
AX~Per  & T60     & UBVR  &    May 2019 &    Oct 2024 \\
        & \asas & bc+bd & 17 Dec 2014 & 20 Nov 2018 \\
        & KWS     & BVIc  & 29 Aug 2011 & 8 Jan 2025  \\
        & S19     & UBVRcIc & Nov 2011 & Dec 2018 \\
        & TESS    & Sector 58  & 29 Oct 2022 & 26 Nov 2022 \\
CI~Cyg  & T60     & UBVR  &    May 2019 &    Oct 2024 \\
        & \asas & bb    & 27 Mar 2015 & 23 Oct 2017 \\
        & KWS     & BVIc  &  5 Jun 2011 & 29 Aug 2023 \\
        & S19     & UBVRcIc & Nov 2011 & Dec 2018 \\
        & TESS    & Sector 74 & 03 Jan 2024 & 30 Jan 2024 \\
        & TESS    & Sector 75 & 30 Jan 2024 & 26 Feb 2024 \\
        & TESS    & Sector 81 & 15 Jul 2024 & 10 Aug 2024 \\
        & TESS    & Sector 82 & 10 Aug 2024 & 05 Sep 2024 \\
Z~And   & T60     & UBVR  &    May 2019 &    Oct 2024 \\
        & \asas & ba    & 23 May 2015 & 26 Nov 2018 \\
        & KWS     & BVIc  &  3 Aug 2011 & 28 Nov 2023 \\
        & S19     & UBVRcIc & Nov 2011 & Dec 2018 \\
        & TESS    & Sector 57  & 30 Sep 2022 & 29 Oct 2022 \\
\hline
\end{tabular}
}
\label{tab:obslog}
\end{table}

\section{Observations} \label{sec:obs}
Multi-wavelength optical photometric data, covering 14~years of observations from both space- and ground-based facilities, have been combined in this study. The complete observing log is presented in Table~\ref{tab:obslog}. 
We present the observations spanning a 14~yr observation period for \axper, \cicyg and \zand in Fig.~\ref{fig:axper_all}, Fig.~\ref{fig:cicyg_all} and Fig.~\ref{fig:zand_all}, respectively.

\subsection{TUG Observations}
We performed optical photometric observations of the objects with 60~cm (T60) robotic telescopes at TÜBİTAK National Observatory (TUG) in Türkiye.
The observational period is between May~2019 and October~2024 by
Bessel \textit{U}, \textit{B}, \textit{V} and \textit{R}-bands. 
The T60 data were reduced and analyzed using AstroImageJ \citep{Collins17}, and differential photometry was performed using comparison stars in the CCD field.

The multicolor T60 photometry is used to construct the color–time diagram.
Because the number of exposures obtained per night differs between bands, the data are nightly averaged in both brightness and time.

\subsection{Observations by \citet{Sekeras19} (S19)}
The objects were also observed 
by multicolor photometry at the Star\'a Lesn\'a Observatory founded by the Astronomical Institute of the Slovak Academy of Sciences, as presented in \citet{Sekeras19} (hereafter S19). 
The observations\footnote{For data and their details cf. \protect\url{https://www.astro.sk/caosp/Eedition/FullTexts/vol49no1/pp19-66.dat/} and \citet{Sekeras19}} were performed using two Cassegrain telescopes with 60~cm and a Maksutov-Cassegrain telescope with 18~cm.
The observational period of the data overlaps with half the period of the KWS data, and the T60 observations almost follow these observations. 
For all objects, observations with photometric uncertainties exceeding 0.5~mag were removed from the datasets.

\subsection{\asas Observations}
All-Sky Automated Survey for Supernovae \citep[\asas;][]{Shappee14,Kochanek17}\footnote{\protect\url{https://asas-sn.osu.edu/}} is an optical sky survey observing by \textit{V}-band. The objects were observed with the ba, bb, bc, and bd cameras at the Hawaii observatory (see Table~\ref{tab:obslog}). The \asas observations do not overlap temporally with the T60 data; however, the T60 monitoring began shortly after the \asas coverage ended.

\subsection{KWS Observations}
Kamogata/Kiso/Kyoto Wide-field Survey \citep[KWS;][]{Maehara14}\footnote{\protect\url{http://kws.cetus-net.org/~maehara/VSdata.py}}
is another program conducting observations in the \textit{B}, \textit{V} and $I_c$-bands. However, only a small amount of $B$-band data is available for the objects, particularly for \axper and \zand. Observation dates in the KWS archive are provided in Gregorian calendar format; therefore, we convert them to Julian dates (JD) using an online converter tool\footnote{\protect\url{https://onlinetools.com/time/convert-date-to-julian-day}}.
The observational period for our objects spans a broader interval than that of the other missions and encompasses the epochs covered by \asas, S19, and T60. Consequently, we present over 13~yr of multi-wavelength optical photometry for all three objects. 

\subsection{TESS Observations}
The Transiting Exoplanet Survey Satellite \citep[TESS;][]{Ricker14}
is a space-based observatory designed to obtain high-precision time-series photometry of individual targets across the entire sky.
The Sector information of the objects is present in Table~\ref{tab:obslog}.
All TESS data were retrieved from the Mikulski Archive for Space Telescopes (MAST) database\footnote{\protect\url{https://mast.stsci.edu/portal/}}. Background-subtracted light curves were generated using the \textit{TESSCut} \citep{Brasseur19}
and \textit{lightkurve}
\textit{Python} \citep{lightkurve18} packages. Postcard images were cropped to 11 × 11 pixels, and standard apertures were applied based on the point-spread function (PSF) results provided by \textit{lightkurve}. Finally, flux time series were extracted for all targets.
The observation windows are shown in the figures of the long-term light variations as vertical red bands. 

TESS observations were acquired with different exposure times depending on the sector. Among the TESS data for our targets, the longest exposure corresponds to Sector~17 of the \zand system, with an exposure time of 1425~s (23.75~min). However, discrepancies were found between the downloaded data and the light curves produced using \textit{TESSCut}. Inspection of the field image revealed that the source was located near the edge of the detector. Consequently, due to the unreliability of the data, observations from this sector were excluded from the analysis. 
Also, in Sector~84 of \zand,  the corresponding light curve is not available in the database, likely because the target lies too close to the CCD edge, although the sector includes an observation of the source.
Similar issues were encountered in some sectors for all three systems. In Table~\ref{tab:obslog}, TESS observations used for this study are presented.

\subsection{Gaia Observations}
In Gaia Data Release 3 (DR3) \citep{GaiaCollab21}, \axper, \cicyg and \zand is identified by ID409028569033785728, ID2047647756007944576 and ID1941894322438077312, respectively. Their distances are estimated using the parallax measurements from Gaia~DR3 and the calculations by \citet{Bailer-jones21}. Due to the variable nature of our sources, the geometric distances ($r_{\rm geo}$) provided in the catalog were adopted instead of the photometric values.
These distances are $2.112\pm0.130~\rm kpc$, 
$1.943\pm0.095~\rm kpc$
$1.920\pm0.079~\rm kpc$
for \axper, \cicyg and \zand, respectively.

\subsection{Frequency analysis of the observations}
We have employed the \textit{Lomb-Scargle periodogram} \citep{Lomb76,Scargle82,Vanderplas18} placed in the package of \textit{astropy}\footnote{\protect\url{https://docs.astropy.org/en/stable/index.html}} to analyze the data of each band for each observatory. The Lomb-Scargle periodogram is a useful tool to process the irregular signals. Time-series analysis can provide more sensitive solutions in the analysis of data sets with irregular interruptions and non-uniform noise.
The method operates by fitting sinusoid-like models to the data in order to compute the power spectrum of the frequencies. In this way, the amplitude distribution of periodic signals at different frequencies is obtained, allowing the identification of the most significant periodic components.

The peak point of the periodical signal can be represented by its probability and the peak height functions placed in \textit{astropy}, defined as \textit{false\_alarm\_probability()} and \textit{false\_alarm\_level()}, respectively. For probability calculations, we adopt the statistical approach suggested by \citet{Baluev08}.

A frequency analysis was performed for each observatory, applied sequentially across all photometric bands, because we have multi-band observations of several observatories. 
Though every mission has different resolutions, the signal strengths are five times higher than the background noise \citep{Scargle82,Horne86,Zechmeister09}, meaning that the frequencies we obtained are significant at the 5$\sigma$ level.
We also present 3$\sigma$ levels in the graphs in the necessary cases.

\section{Results} \label{sec:results}

\subsection{AX~Per}

\begin{figure*}[htb!]
    \centering
    \includegraphics[width=0.99\linewidth]{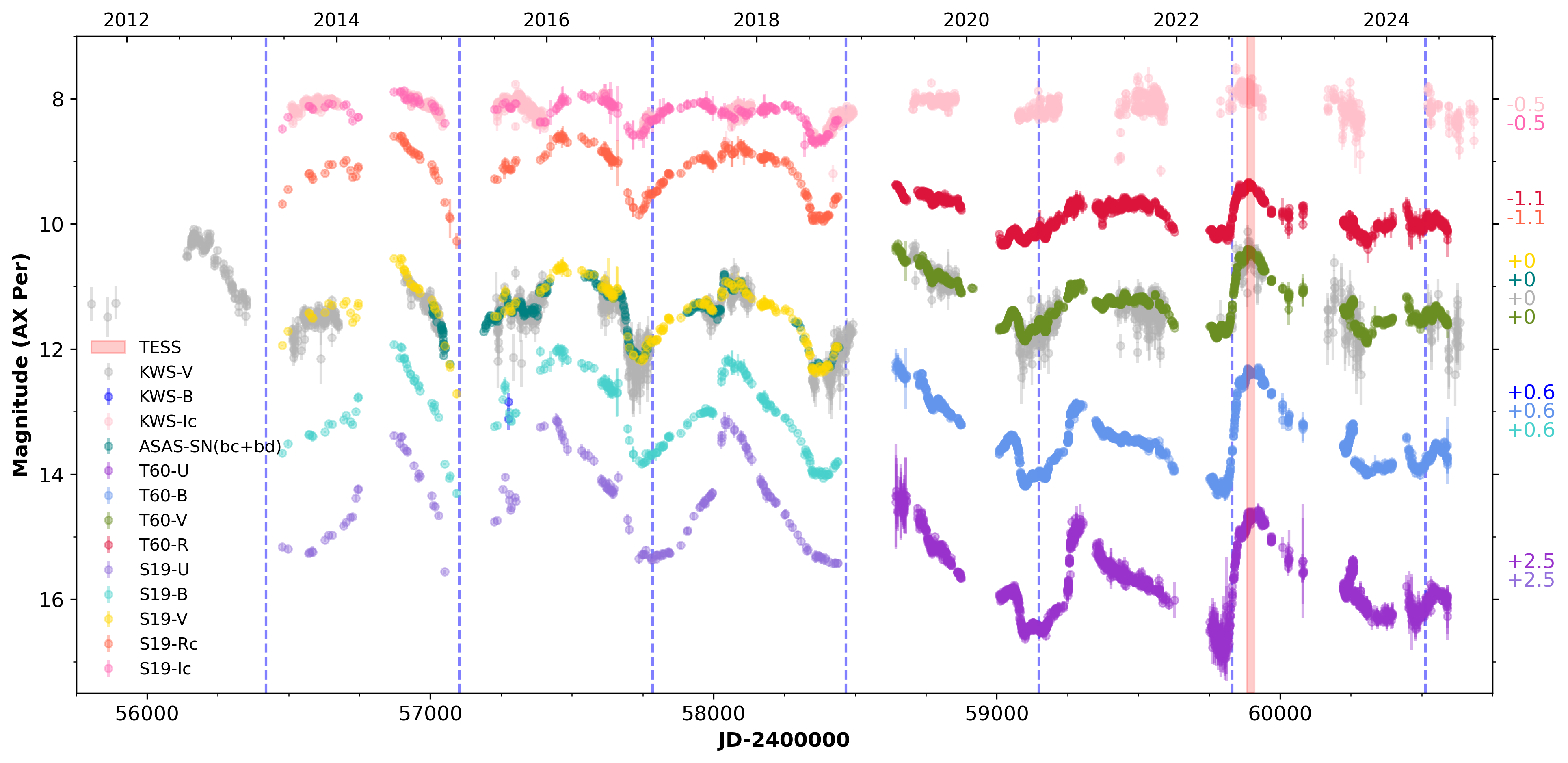}
    \caption{The light variations of all-bands data from \asas, KWS, S19 and T60 for \axper. 
    The red vertical bands span the TESS observation Sector(s), and the blue dashed lines correspond to the ephemeris of the inferior spectroscopic conjunction of the red giant according to \citet{Fekel00_II}.
    The offset values with the same color codes are on the outside of the graph.}
    \label{fig:axper_all}
\end{figure*}
\subsubsection{Long-term light curve }
A 14~yr observational period of \axper is presented in~Fig.\ref{fig:axper_all}. \axper is in an active phase during the entire monitoring period. Throughout this interval, the system exhibits high-amplitude outbursts, 
with \textit{V}-band brightness increases of up to approximately 10~mag (see Fig.\ref{fig:axper_all}). 
Three distinct brightness maxima brighter than 10.5~mag have been recorded. Deep and narrow minima are observed during this phase; however, after JD~$\approx$~2459000 the eclipses become shallower, with increasing base brightness levels and broader minima showing low-amplitude wave-like behaviors.

\begin{figure}
\centering
\includegraphics[width=0.99\linewidth]{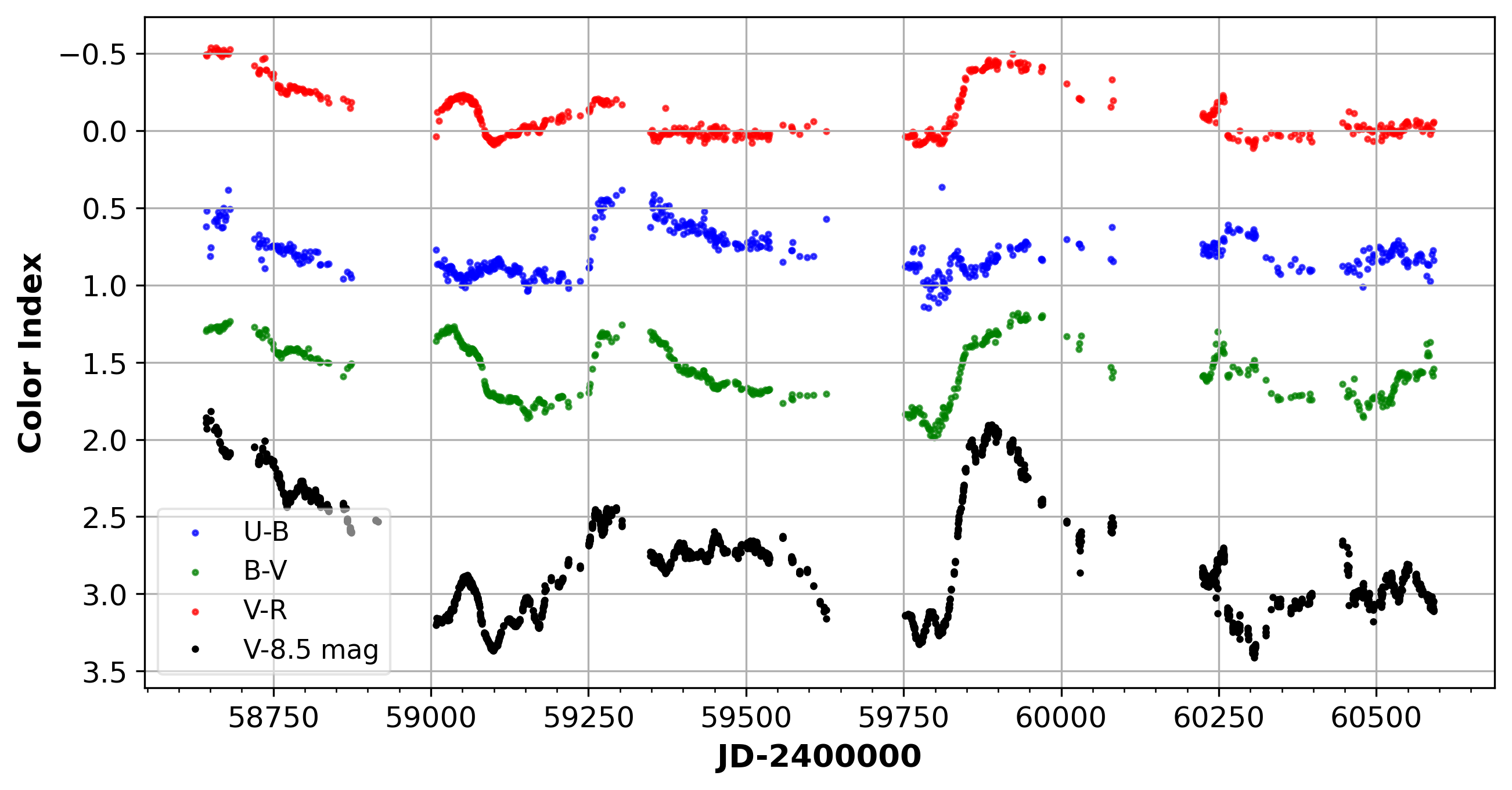}
\caption{TUG T60 $U-B$, $B-V$ and $V-R$ colors for \axper. T60 \textit{V}-band is also shown by offset by -8.5~mag for easy comparisons with the colors.}
\label{fig:t60_axper_colordiagram}
\end{figure}

\textit{Color-time diagram:}
At the beginning of the T60 observations, the system exhibits light contributions from both companions due to its orbital position. The light curve (Fig.~\ref{fig:t60_axper_colordiagram}) starts shortly after the maximum, showing a gradual decline in brightness. This period is associated with the system's active phase. Between JD~$\approx 2458640 - 2458880$, the system’s total luminosity decreases while its colors become progressively redder. 
This trend may indicate a diminishing contribution from the hot component and the associated ionized regions, particularly reflected in the decrease of the \textit{U}-band flux.

After a gap in the observations, data starting around JD~$\approx$~2459000 show bluer \textit{U–B} and \textit{B–V} colors. This interval corresponds roughly to the system’s eclipse minimum (indicated by the blue dashed lines in Fig.~\ref{fig:axper_all}). Within JD~$\approx2458745 - 2459100$, a distinct hump-like brightening is detected, with the strongest amplitude in the \textit{U} and \textit{B} bands. 
The presence of this brightening despite the potential eclipse of the hot star suggests that a significant portion of the emission originates from the extended ionized nebula.

Following this phase, a sequence of small-amplitude, wave-like periodic variations is observed between JD~$\approx2459100 - 2459570$ (the periodical variation is \simi37 and \simi75~d). 
These brightenings show an inverse relationship between brightness and color, while the \textit{V}-band brightens (or fades), the colors become redder (or bluer). 
Such behavior is observationally consistent with the variations typically originating from the cool giant, possibly due to pulsations, as also supported by similar fluctuations in the \textit{R}-band.
Comparable low-amplitude variations are also detected during JD~$\approx2459550 - 2460100$ and JD~$\approx2460460 - 2460550$.

At JD~$\approx2459570$, a noticeable fading is observed simultaneously in all filters, while the color indices remain nearly constant, consistent with the onset of an eclipse event.
Around JD~$\approx2460090$, a prominent brightening event dominates across all bands, with the strongest variations seen in the \textit{U} and \textit{B} filters. Between JD~$\approx2459820$ and 2460090, the variations in these bluer bands are relatively small compared to other filters, suggesting that the outburst mainly manifests in the \textit{V}-band. 
At the peak of the outburst, there were slight increases in both \textit{U–B} and \textit{B–V} values. 
Subsequently, the outburst declines, and the system begins to fade. During this fading phase, \textit{U–B} remains almost constant,  with the brightness decreases without a significant change.
This roughly 250~d long event is consistent with the characteristics of a \textit{combination nova-type outburst}.

Around JD~$\approx246050$, a small-amplitude brightening is detected, followed by a sharp drop in brightness. This likely corresponds to a short outburst occurring just before the eclipse of the hot component, during which the brightening source begins to be obscured by the red giant (JD~$\approx2460260$).

\begin{figure}
\centering
\includegraphics[width=\linewidth]{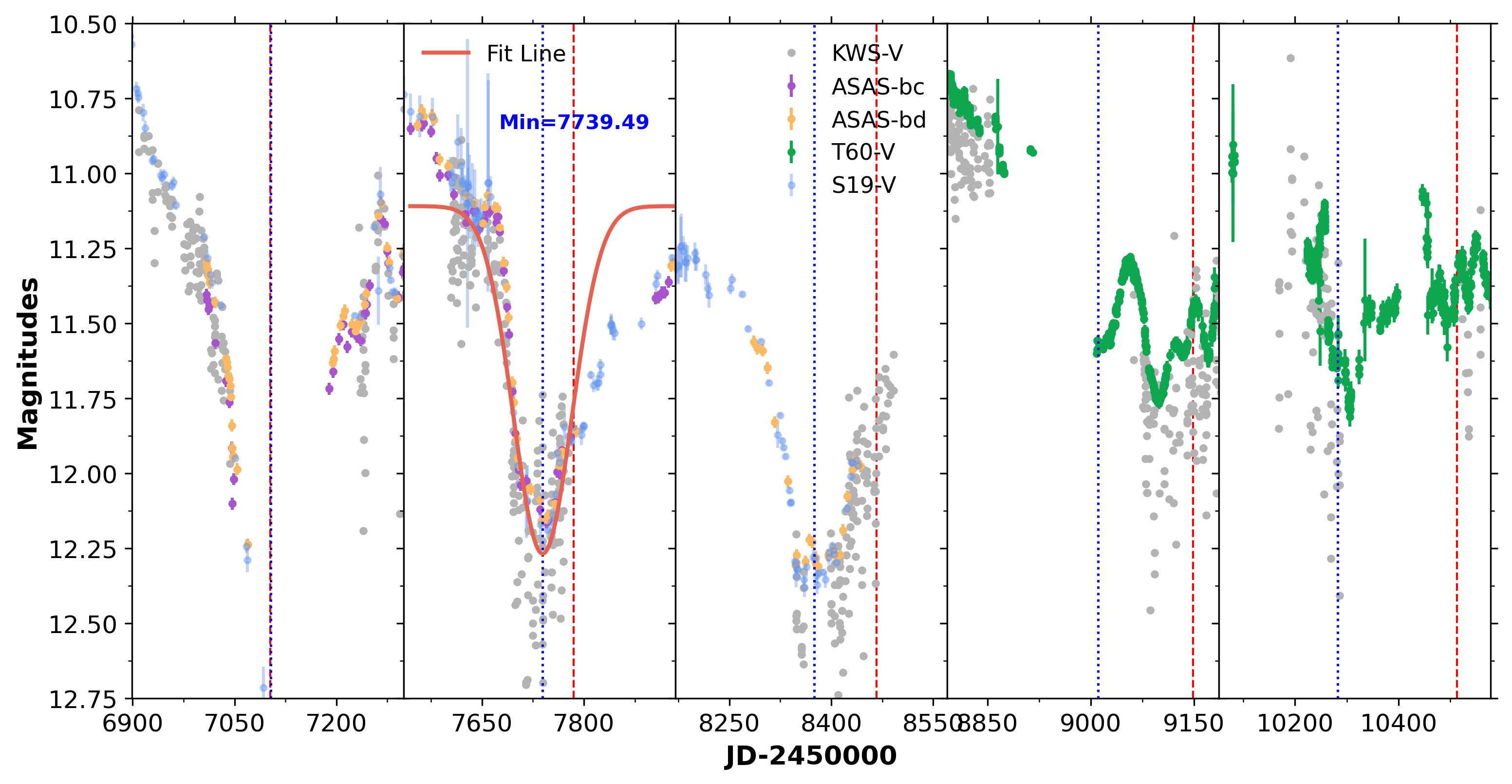}
\caption{The expected minima locations for \axper. The blue dotted lines indicate the positions of the observed minima based on the fit line, while the red dashed lines correspond to their positions calculated according to the ephemeris of \citet{Fekel00_II}. }
\label{fig:axper_minima}
\end{figure}

\textit{Eclipsing Minima:}
During the observational period spanning more than 13 years, the system has reached its minimum brightness approximately 8 times. However, due to insufficient data coverage at certain epochs, only five intervals can be considered for the analysis of eclipse minima.
This time interval to examine the ephemeris of the orbitally related minima is inefficient and is additionally influenced by the intrinsic activity of the system. Rather than attempting to define the true orbital period, we here show how the lower limit of the observed minima varies and is influenced by the system’s activity, and compare our observed minima with the predicted dates from the spectroscopic ephemeris.

For the time interval with the densest coverage, a Gaussian fit was applied to determine the minimum, yielding $\rm JD = 2457739.49$ (see the second panel from the left in Fig.~\ref{fig:axper_minima}).
The fitting was performed over the range $2457545 < \rm JD < 2457931$ using the combined KWS, S19, and \asas datasets.
While the photometric uncertainties from KWS were included in the numerical analysis, they were omitted from the plotted figure for clarity. Using this minimum as a reference point and considering the 635.76~d periodicity derived from the \textit{V}-band periodogram, the epochs of other eclipse minima were estimated. 
Afterward, we predicted dates from the spectroscopic ephemeris from \citet[$\rm JD_{spec}=2450963.8(\pm5.6)+682.1(\pm1.4)\times \rm E$;][]{Fekel00_II} and presented them with the red dashed lines on Fig~\ref{fig:axper_minima}.
This figure shows that determining the precise timing of the minima is particularly challenging, since \axper exhibits outburst-like structures even during eclipse phases.

The wave-like fluctuations observed within the minima (such as those detected during the T60 observing intervals) further complicate these determinations. 
During quiescent phases, the dominant optical emission associated with the hot component and the nebula does not necessarily coincide with the center of mass of the hot star, while during strong outbursts, the region of enhanced emission may be closer to the hot component. As the system transitions between quiescent and active states, the dominant source of optical radiation shifts due to changes in the relative contributions of the hot companion and the surrounding nebula.
This redistribution of the light-emitting regions leads to apparent displacements of the photometric minima, which do not reflect true variations in the orbital period but rather changes in the location of the system’s light center \citep{Skopal98}. 
This distinction helps to separate the stable orbital geometry of the system from the apparent shifts of the light centers that occur during active phases.

\subsubsection{TESS observations}
Photometric short-term variability of the three target systems was investigated using data obtained with the TESS. Only a single TESS sector is available for the \axper system (Sector~57). The observation window spans $2459882 < \rm BJD < 2459910$. The TESS observations of \axper are located at a phase where both components of the system are visible, and the brightness is at its post-eclipse maximum (see Fig.~\ref{fig:axper_all}). The light curves used in this data correspond to the 120~s short-cadence mode, allowing us to probe variability on minute-level timescales with high photometric precision.

\begin{figure}
\centering
\includegraphics[width=\linewidth]{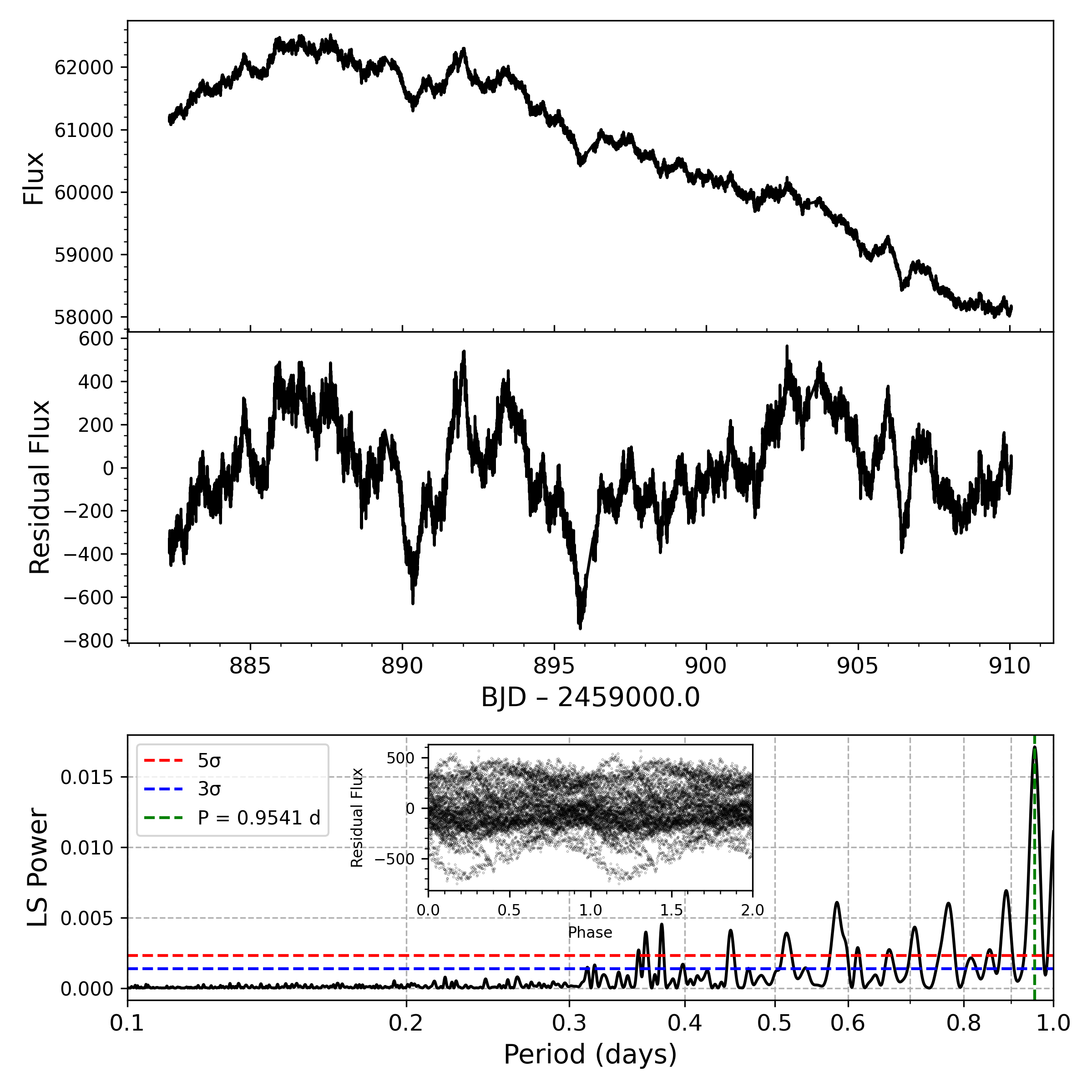}
\caption{TESS observations and Lomb-Scargle Periodogram of \axper. From top to bottom, the raw TESS light curve, the detrended light curve, and the periodogram of the detrended light curve with phase-folded light curve in the mini panel, respectively. At the bottom panel, the vertical green dashed line marks the period with the highest amplitude, shown correspond to the period of 
$P = 0.9541\,\mathrm{d}$ ($f = 1.048\,\mathrm{d^{-1}}$).}
\label{fig:axper_tess}
\end{figure}
As seen in Fig.~\ref{fig:axper_tess}, throughout the sector, the flux exhibits a slight rise at the beginning of the observations, followed by a gradual decline. The light curve also displays short-term, flickering-like fluctuations (a photometric amplitude of about 200–400~electron~s$^{-1}$ in PDCSAP flux unit), a behavior not observed in the TESS data of the other systems. 
When the long-term light curve of the system is compared with the TESS timing window, 
it becomes evident that the TESS observed from near the peak of a broad till to the end of the hill-shaped maximum.
To enhance the detectability of short-term variations in the periodogram and to remove the sector-long decreasing trend, the raw TESS light curve is fitted with a second-order polynomial across the sector, and this polynomial trend is subsequently subtracted from the data. 
The resulting detrended light curve is then analyzed using the Lomb–Scargle periodogram. Here, we are interested in investigating the dominant frequencies in the range between 0.01 and 1~days.

\subsubsection{Frequency analysis}
\begin{figure}
\centering
\includegraphics[width=\linewidth]{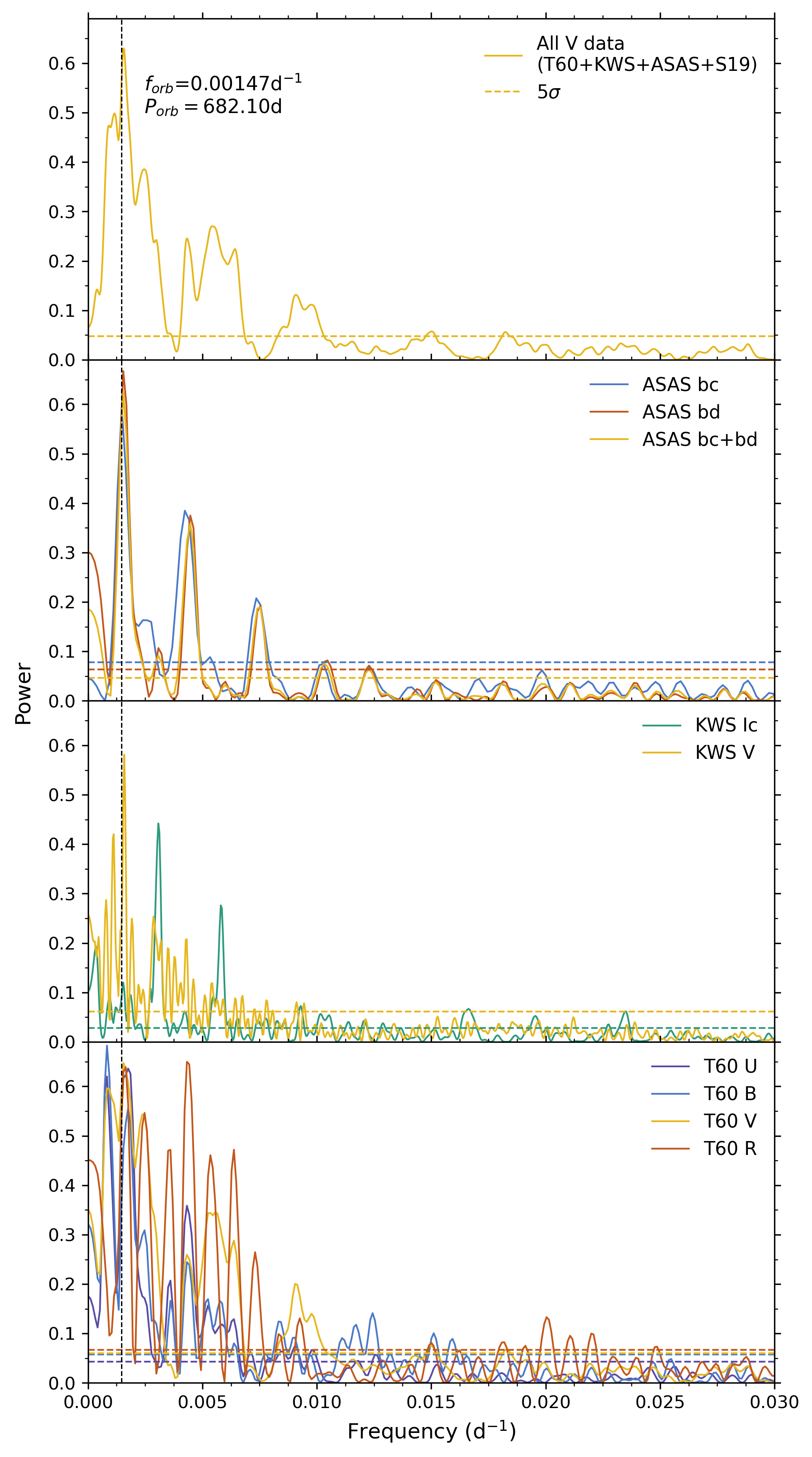}
\caption{Lomb-Scargle Periodogram of \axper for all observations. 
The vertical dashed line denotes the orbital frequency ($f_{\rm orb}$) based on the spectroscopic period of \citet{Fekel00_II}. The horizontal dashed lines, according to the color codes, indicate $5\sigma$ S/N levels of corresponding observations. 
}
\label{fig:axper_freq}
\end{figure}

The light curves of \axper show the wave-like modulations following the main-brightness level with narrow amplitudes and small strengths of magnitudes (clearly seen in T60 observations). We investigate these variations and their characteristic parameters by frequency analysis. Additionally, we present all periodical frequencies obtained from the analysis (see Fig.~\ref{fig:axper_freq}).

\textit{T60 data.} A significant modulation with a period of \simi1213~d ($f = 0.00082 $\rday) is detected in the \textit{U}, \textit{B}, and \textit{V}-bands, although it is not the dominant signal. No corresponding variability is found in the \textit{R} band, whereas in the \textit{B}-band this frequency represents the strongest peak. The trend likely arises from the gradual decline in brightness observed over the monitoring interval, as the system appears brighter at the beginning of the \textit{U} and \textit{B}-band observations. Consequently, the closest frequency to the orbital period in these filters is found near 570~d ($f = 0.00175 $\rday). The 1213~d signal affects the frequency analysis by distorting the dominant peak toward the orbital timescale. To correct for this, the long-term variation is modeled with a sinusoidal function and subtracted from the data before recomputing the periodogram. After detrending, the photometric periodicities of 646.97~d ($f = 0.00155 $\rday) in the \textit{U} band and 648.98~d ($f = 0.00154 $\rday) in the \textit{B}-band are recovered. 
In the \textit{V}-band, a dominant frequency corresponding to a period of \simi~1216~d is also detected. However, removing this long-term variation does not affect the periodicity associated with the orbital cycle. Consequently, both the \textit{V} and \textit{R}-bands exhibit a consistent photometric periodicity of 648.98~d ($f = 0.00154 $\rday).
As noted in the previous section, determining the true orbital period of such systems from photometry alone is inherently challenging. In this study, rather than seeking the true orbital period, we aim to examine how the ephemeris derived from spectroscopic studies compares with the photometric minima obtained from our observations. The orbital period of 682.1 days reported from spectroscopic analyses of \citet{Fekel00_II} is approximately 40 days longer than the photometric period inferred here. As shown in Fig.~\ref{fig:axper_minima}, transitions between quiescent and active phases, together with wave-like fluctuations around the base of the minima, can naturally lead to discrepancies in the observed positions of the photometric minima. 
During quiescent phases, the dominant contribution to the optical emission arises from the extended ionized nebula, whose effective light center does not coincide with the center of mass of the hot component. During active phases, the expansion of the hot component’s photosphere migrates the dominant region of optical emission closer to the hot component \citep{Skopal98}.

\begin{figure}
\centering
\includegraphics[width=\linewidth]{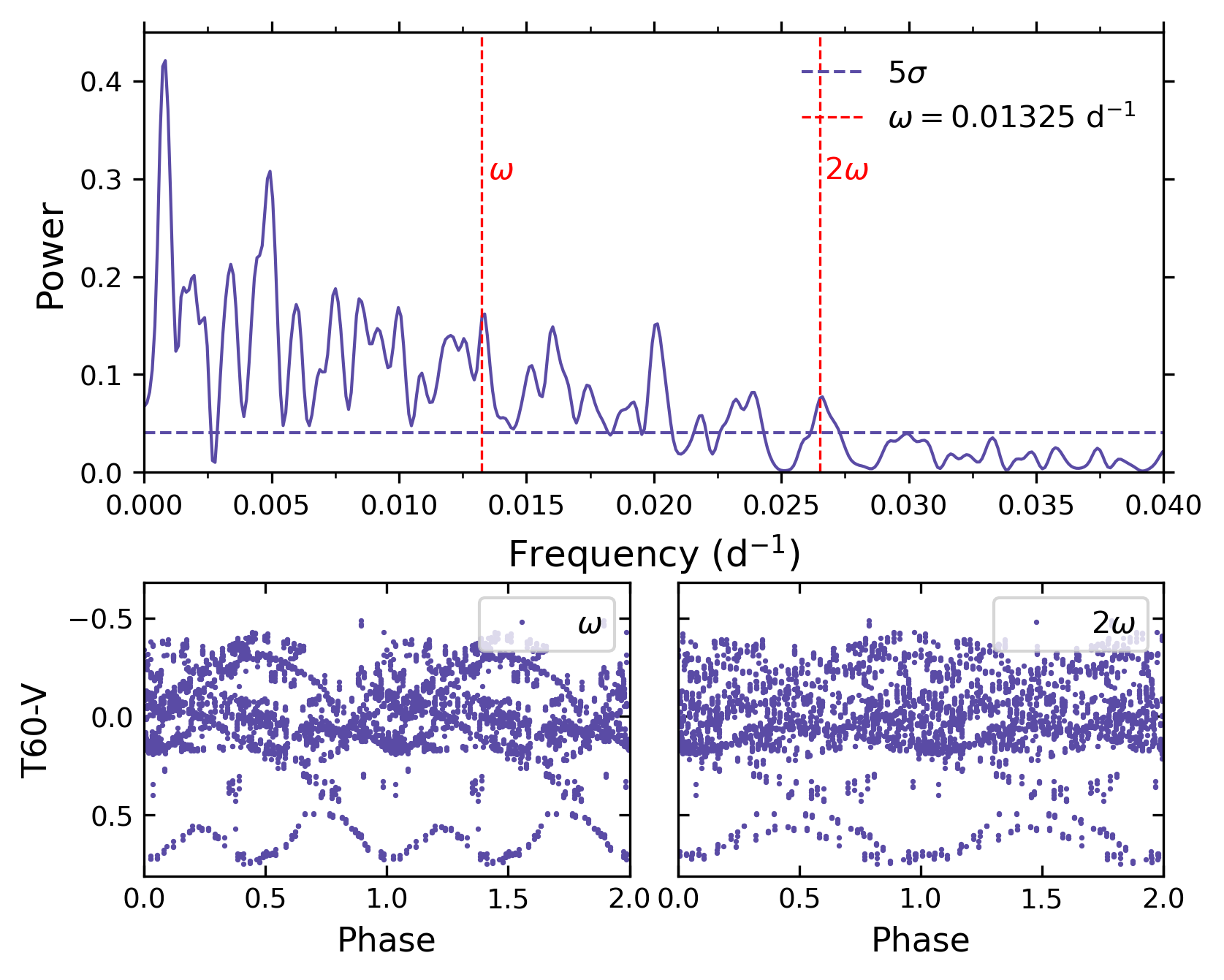}
\caption{Upper panel: Power spectrum of the residual data after removing long-term sinusoidal variations. Lower Panel: Folded-phase diagrams of $\omega$ and $2\omega$ from left to right, respectively.}
\label{fig:axper_p37and75}
\end{figure}

A dominant frequency around 230~d ($f = 0.0043 $\rday) is detected in all photometric bands. This variation likely arises from the nearly periodic recurrence of outbursts and brightness peaks in the light curve, which occur on timescales of roughly 200–230~d. To better resolve lower-amplitude variations, the long-term modulations with periods of 1216 and 648.98~d are modeled with sinusoidal functions and subtracted from the dataset. After this correction, an “intermediate timescale” variation was identified, with periods of 75.82 ($f = 0.01319 $\rday), 76.05 ($f = 0.01315 $\rday), 74.88 ($f = 0.01335 $\rday), and 74.31~d ($f = 0.01346 $\rday) in the \textit{U}, \textit{B}, \textit{V}, and \textit{R}-bands, respectively. In addition, a secondary frequency corresponding to half of this period (\simi37.44~days; $f = 0.02671 $\rday) is also detected. The resulting periodogram and phase-folded light curves for these frequencies are shown in Fig.~\ref{fig:axper_p37and75}. The small-amplitude variations obtained for the \textit{V} band correspond to periods of 37.73~d ($f = 0.02650 $\rday) and 75.46~d ($f = 0.01325 $\rday), which are in exact resonance with each other.

\textit{KWS data.} The dominant frequency corresponding to the 1213~d variation detected in the T60 observations is also present in the KWS data. For the \textit{V}-band, this strong pre-orbital signal corresponds to a period of 1271.49~d. The orbital period is determined to be 635.76~days ($f = 0.0016 $\rday) (see in Fig~\ref{fig:axper_freq}). The error sensitivity of the KWS data is lower than that of the other observatories. Meanwhile, this scattered dataset does not provide reliable results for detecting variations on the order of several tens of days. Nevertheless, after removing the higher-amplitude variations (1271.49 and 635.76~d) from the data, a low-amplitude periodicity of \simi78.95~d ($f = 0.0127 $\rday) is identified in the \textit{V}-band.

KWS observations are obtained in separate observing intervals, resulting in distinct clusters of data. For the \textit{V}-band, these clusters typically span 170–200~d, with gaps between them lasting about 200–230 days. During the frequency analysis, alias frequencies corresponding to these time intervals are identified. In the process of subtracting the dominant variations to isolate weaker signals, an additional modulation with a period of 210.08~d ($f = 0.00476 $\rday) is detected, which is interpreted as an alias frequency.
In the \textit{I\textsubscript{c}}-band, the dominant frequency corresponds to a period of 321.2~d ($f = 0.0031 $\rday), close to a resonance with the orbital frequency. The second strongest signal appears at 171.13~d ($f = 0.0058 $\rday). It is important to note that KWS \textit{I\textsubscript{c}}-band data are conducted in discrete time segments, each lasting approximately 170~d, indicating that this frequency likely arises from the observing cadence and represents an alias.
After the dominant variation is modeled with a sinusoidal function and subtracted from the dataset, a prominent frequency near the orbital period is identified, corresponding to 632.67~d ($f = 0.00158 $\rday). This variation is subsequently represented by a sinusoidal fit and removed, revealing an intermediate-timescale modulation with a period of 77.33~d ($f = 0.0129 $\rday).

\textit{\asas data.} \axper has been monitored by \asas over a time span of \simi1434~d (3.93~yr). In the frequency analysis, data from all cameras is combined (\textit{bc+bd}) to be analyzed together, after which the datasets from each camera are examined separately. The most prominent frequency detected in all cases corresponds to the orbital timescale, with periods of 651.83~d ($f = 0.00153 $\rday) for \textit{bc+bd}, 680.66~d ($f = 0.00147 $\rday) for \textit{bc}, and 651.82~d ($f = 0.00153 $\rday) for \textit{bd}. The slight differences between cameras arise from the distinct durations of their observing baselines and, consequently, the number of covered orbital cycles. The \textit{bc} camera, with a total monitoring span of about 1089~d, covers 2 eclipsing minima, whereas the \textit{bd} camera, spanning 1434~d, samples 3 minima over \simi2.2 orbital cycles.

\textit{All monitoring data.} Additionally, we combine all \textit{V}-band data of T60, KWS, and \asas to increase in data points. The predicted orbital period is obtained 635.76~d ($f = 0.0016 $\rday). After this variation is modeled with a sinusoidal function and subtracted from the dataset, an intermediate-timescale variation is measured as 75.26~d ($f = 0.0133 $\rday). 

\textit{TESS data.} We perform a period analysis of the TESS Sector~58 light curve of \axper. A third-degree polynomial is fitted to the full dataset to remove long-term trends, and the analysis is carried out on the resulting detrended light curve. The highest-amplitude power in the periodogram lies between approximately 0.3 and 1~d. Periodicities shorter than 0.1~d fall well below the $3\sigma$ significance threshold and were therefore excluded from further consideration. The most significant signal detected is a period of \simi0.95~d (22.9~h), which may correspond to low-amplitude oscillations present in the light curve and is illustrated in Fig.~\ref{fig:axper_tess}. The phase-folded light curve exhibits a sinusoidal modulation. No shorter-period variability that could be associated with the white dwarf spin was identified.

\begin{figure*}
    \centering
    \includegraphics[width=\linewidth]{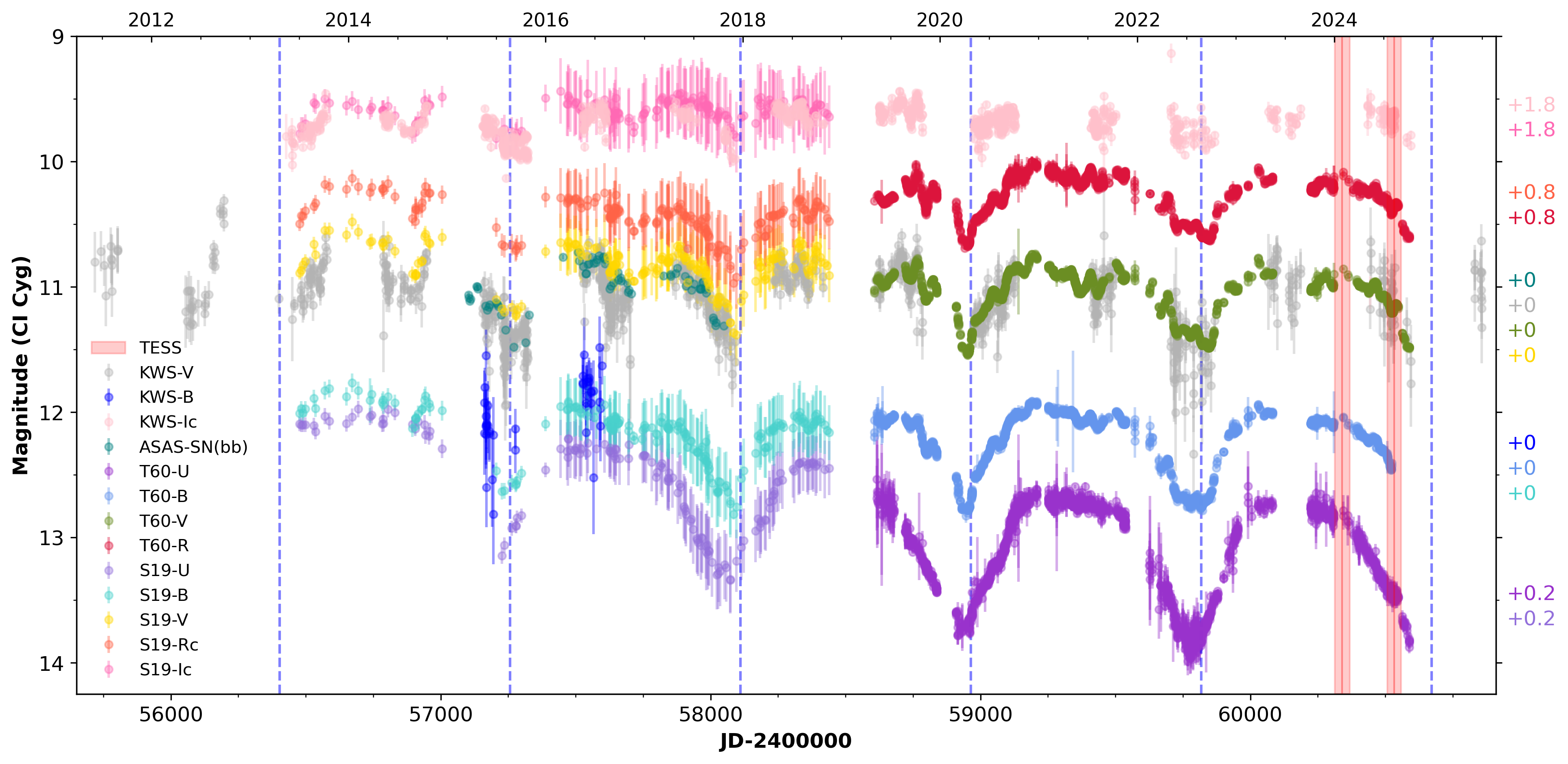}
    \caption{The light variations of all-bands data from \asas, KWS, S19 and T60 for \cicyg. The red vertical bands span the TESS observation Sector(s), and the blue dashed lines correspond to the ephemeris referring to the inferior spectroscopic conjunction of the red giant according to \citet{Fekel00_I}. The offset values with the same color codes are on the outside of the graph.}
    \label{fig:cicyg_all}
\end{figure*}
\subsection{CI~Cyg}

\subsubsection{Long-term light curve }
During \simi14~yr observing interval analyzed, the system of \cicyg predominantly exhibits quiescent-phase behavior (see Fig.~\ref{fig:cicyg_all}). Within this interval, only a single pattern is detected in which the system’s brightness exceeds 10.5~mag (JD\simi2456150). In fact, this indicates that the quiescent phase essentially begins following this outburst. Over the light variation, the system is characterized by low-energy, short-lived brightness fluctuations rather than sudden, high-energy outbursts throughout its quiescent state.

\textit{T60 data.}
\cicyg was monitored with the TUG~T60 telescope over a period of approximately 5.4~yr, covering about 2.3 orbital cycles. 
The light variations display clear wave-like fluctuations with low amplitudes, particularly in the \textit{B}, \textit{V}, and \textit{R}-bands, where such modulations are evident along both the ingress and egress branches of the eclipses. 
To minimize the influence of such variations when determining the primary minima, the \textit{U}-band data (being the least affected by these fluctuations) are used.
Gaussian fits are applied to \simi600~d data segments to estimate the times of minima. 
The resulting minima are $\rm JD_{minI}=2458921.5689$ and $\rm JD_{minII}=2459779.6926$, corresponding to a time interval of \simi858.1237~d. Accordingly, the T60 photometric ephemeris is expressed as
$\rm JD_{T60}=2458921.5689 + 858.1237 \times \rm E$.

\begin{table}
    \centering
    \caption{The eclipse minima of CI Cyg determined from Gaussian fits to the T60 observations.}
    \begin{tabular}{ccc}
    \hline
    \hline
    \bf Bands & \bf 1. Primary Min. & \bf 2. Primary Min.\\ 
    \hline
    \textit{U} & 2458921.5689 & 2459779.6926 \\ 
    \textit{B} & 2458928.0150 & 2459787.0632 \\
    \textit{V} & 2458939.7697 & 2459802.2821 \\ 
    \textit{R} & 2458937.2872 & 2459804.9469 \\
    \hline
    \end{tabular}
    \label{tab:cicyg_minima}
\end{table}
The ephemeris derived by \citet{Skopal12} from eclipse observations is given as $\rm JD_{ecl.}= (2441838.8\pm1.3) + (852.98\pm0.15) \times \rm E$.
There is a difference of 20.03 orbital cycles between the reference epoch ($T_0$) reported by \citet{Skopal12} and the first minimum obtained from the T60 observations, corresponding to a phase shift of approximately 23.17~d. Considering the orbital period of 852.98~d, this offset represents a deviation of about 2.7\%. The primary minima determined for all photometric filters are listed in Tab.~\ref{tab:cicyg_minima}.

\textit{Color-time diagram.}
During the T60 monitoring, \cicyg remain in a quiescent state; therefore, no sudden or intense outbursts capable of dominating the overall light variability are detected. The primary variability in \cicyg is conventionally the orbital motion, while a secondary contribution arises from low-amplitude, wave-like brightness modulations with a characteristic period, similar to those identified in \axper. 

\begin{figure}
    \centering
    \includegraphics[width=0.99\linewidth]{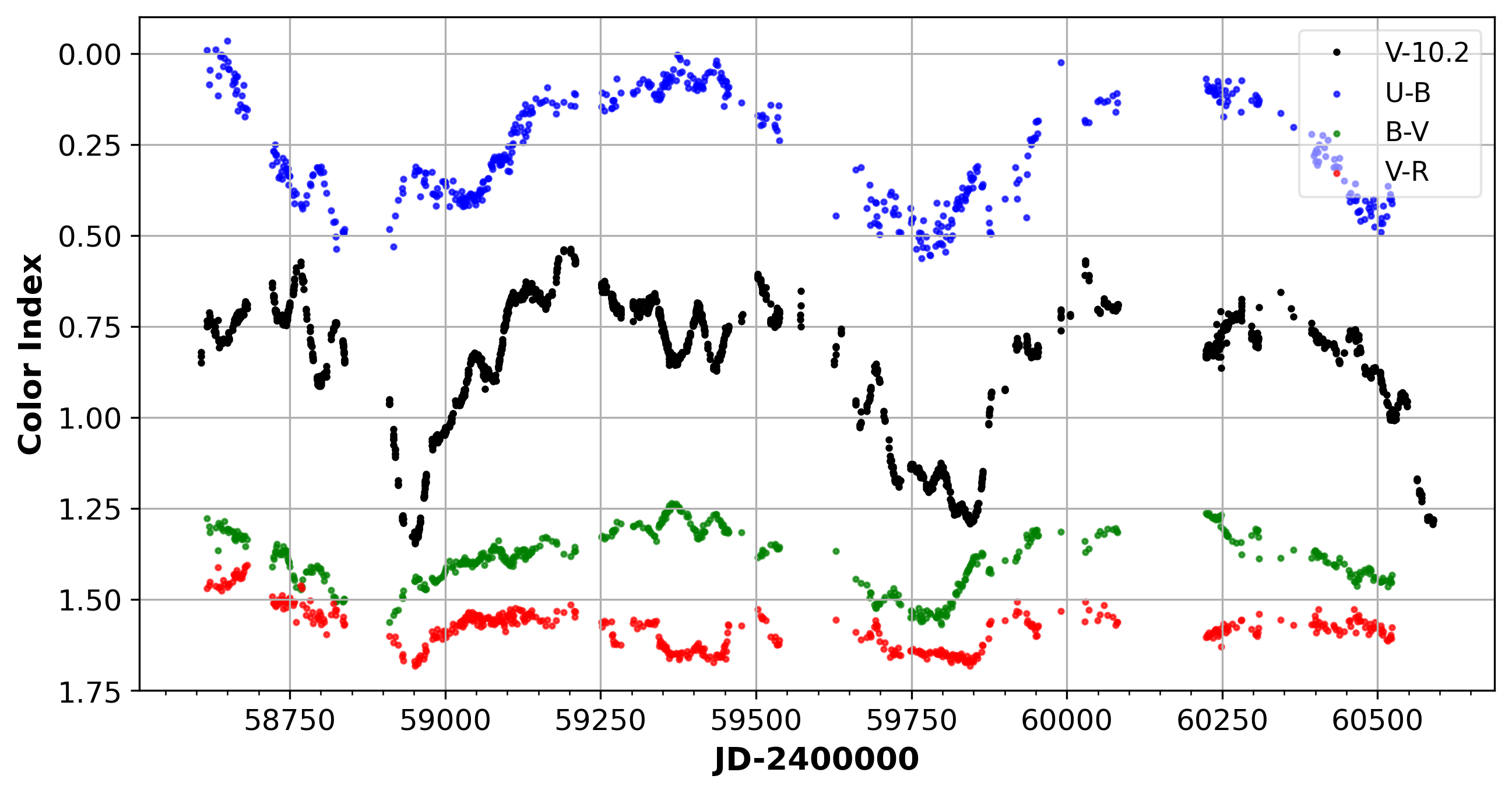}
    \caption{TUG T60 $U-B$, $B-V$ and $V-R$ colors for \cicyg. T60 \textit{V}-band is also shown by offset by -10.2~mag for easy comparisons with the colors.}
    \label{fig:cicyg_color}
\end{figure}
Fig.~\ref{fig:cicyg_color} presents the color diagram constructed from T60 multi-band photometric observations. The light curves display wave-like modulations that appear to vary with a characteristic timescale (see Sect.~\ref{sec:cicyg_freq}). 
While the \textit{V}-band variability closely follows the behavior of the \textit{V–R} color (with maxima and minima occurring simultaneously), the maxima in the \textit{V}-band correspond to decreases in the \textit{U–B} and \textit{B–V} color indices, indicating an inverse color-magnitude relation.
The fact that different color indices do not vary coherently suggests that these wave-like modulations cannot be straightforwardly attributed to a single source of variability. In particular, variations dominated purely by pulsations of the cool giant would be expected to produce more uniform color behavior.
The observed color-dependent response may therefore reflect the blended contributions of the three basic radiation components in symbiotic systems (the cool giant, the hot star, and the ionized nebula), which are mixed within the photometric filters \citep{Muerset91, Skopal05}. However, photometric data alone do not allow these contributions to be uniquely disentangled, and the underlying physical mechanism cannot be unambiguously identified on this basis.

Another noteworthy feature is the presence of wave-like modulations even during the eclipse of the hot component. While these variations are visible across all bands during the eclipse, the colors (particularly \textit{V–R}) remain nearly constant. This suggests that the wave-like modulations are not directly associated with the hot component yet may instead arise from quasi-periodic pulsations of the red giant. The fact that these oscillations are most prominent in the \textit{V} and \textit{R}-bands, with only minor color changes, implies that the pulsation primarily modulates the system’s overall brightness. The wave patterns are especially pronounced in the \textit{V} and \textit{R}-bands, indicating that a quasi-periodic pulsation of the red giant is a likely origin. However, the lack of synchronicity among all color indices hints at a more complex process involving both pulsation and disk effects. Small-amplitude, quasi-periodic pulsations of the red giant could produce the observed modulation in the \textit{V} and \textit{R}-bands, whereas the extended accretion disk of \cicyg \citep{Kenyon91} may influence the hot-component radiation, leading to the inverse color behavior observed in \textit{U–B} and \textit{B–V}.

\subsubsection{TESS observation}
\begin{figure}
\centering
\includegraphics[width=\linewidth]{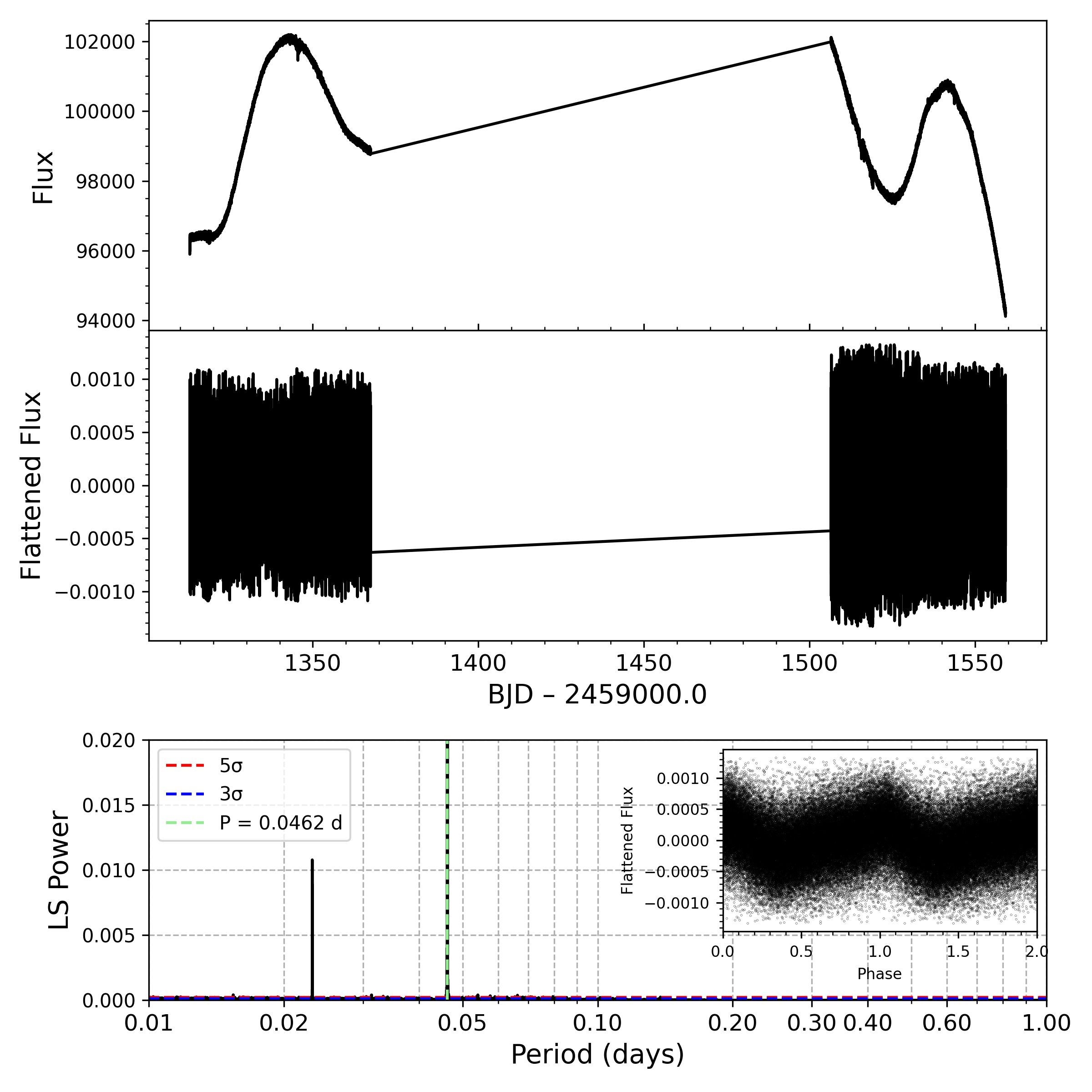}
\caption{TESS data and Lomb-Scargle Periodogram of \cicyg. From top to bottom, the raw TESS light curve, the flattened light curve (Savitzky-Golay filter), and the periodogram of the flattened light curve with phase-folded light curve in the mini panel, respectively. 
At the bottom panel, the vertical red dashed line marks the period with the highest amplitude, and the two peaks shown correspond to periods of 
$P = 0.0462\,\mathrm{d}$ ($f = 21.65\,\mathrm{d^{-1}}$) and its first harmonic 
$P/2 = 0.0231\,\mathrm{d}$ ($f = 43.29\,\mathrm{d^{-1}}$), right to left, respectively. 
}
\label{fig:cicyg_tess}
\end{figure}

\cicyg is among the systems with the largest number of TESS observations, with data available from four separate TESS sectors. These observations correspond to Sector~74 ($2460312<\rm BJD<2460339$), Sector~75 ($2460339<\rm BJD<2460367$), Sector~81 ($2460506<\rm BJD<2460533$), and Sector~82 ($2460533<\rm BJD<2460559$). As shown in Fig.~\ref{fig:cicyg_all}, the observations are distributed across the long-term light curve. The overall behavior of the long-term light curve is characterized by a decline from a peak-like maximum, superimposed with sinusoidal-like modulations. The behavior seen in the TESS light curves is not uniform when all sectors are combined into a single dataset. The first two sectors (Sectors~74 and 75) exhibit a Gaussian-like profile when merged, whereas the last two sectors (Sectors~81 and 82) display a more sinusoidal pattern. A second-order polynomial representation is applied to the four-sector light curve, bringing all segments to a consistent baseline (see Fig.~\ref{fig:cicyg_tess}). To identify both short- and long-term variations in the light curves, the Lomb–Scargle periodogram analysis is performed on all four sectors, as mentioned in the following section.

\begin{figure}
\centering
\includegraphics[width=\linewidth]{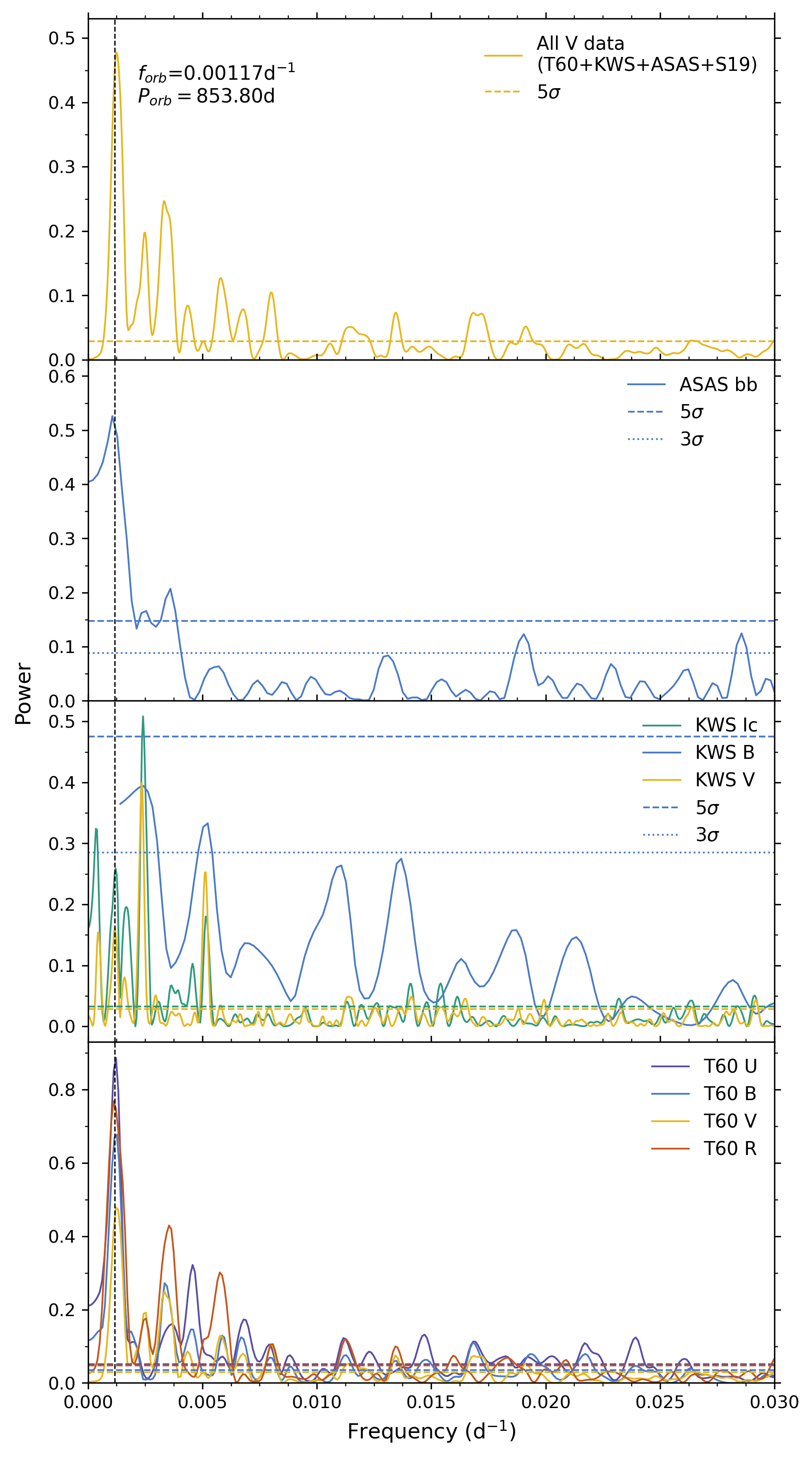}
\caption{Lomb-Scargle Periodogram of \cicyg for all observations. 
The vertical dashed line denotes the orbital frequency ($f_{\rm orb}$) based on the spectroscopic period of \citet{Fekel00_I}. The horizontal dashed and dotted lines according to the color codes indicate $5\sigma$ and $3\sigma$ significances of corresponding observations, respectively. 
}
\label{fig:cicyg_freq}
\end{figure}
\subsubsection{Frequency analysis} \label{sec:cicyg_freq}
The frequency analysis shows that, across all filters, the dominant peak lies close to the orbital periodicity. Fig.~\ref{fig:cicyg_freq} presents the power–frequency spectra derived from multiband photometry across all missions, with the dotted vertical line marking the frequency of the orbital period. The horizontal dashed lines in the power spectrum, color-coded by filter, indicate the $5\sigma$ significance threshold adopted for the detection of periodic signals.

\textit{T60 data.}
For the periodogram derived from the \textit{U}-band of T60, the orbital period is measured as approximately 857.73 days ($f= 0.00117 $\rday). 
The period inferred from the power spectrum (857.73~d) is in excellent agreement with the 858.0287~d interval measured between the two eclipse minima in \textit{U}-band. 

After subtracting the large-amplitude variations from the T60 light curves using sinusoidal fits, a second frequency analysis was performed. The resulting dominant signals occur at $f = 0.01467 $\rday (68.17~d) in the \textit{U} band, $f = 0.01342 $\rday (74.50~d) in both the \textit{B} and \textit{V}-bands, and $f = 0.01344 $\rday (74.39~d) in the \textit{R}-band.

\begin{figure}
\centering
\includegraphics[width=\linewidth]{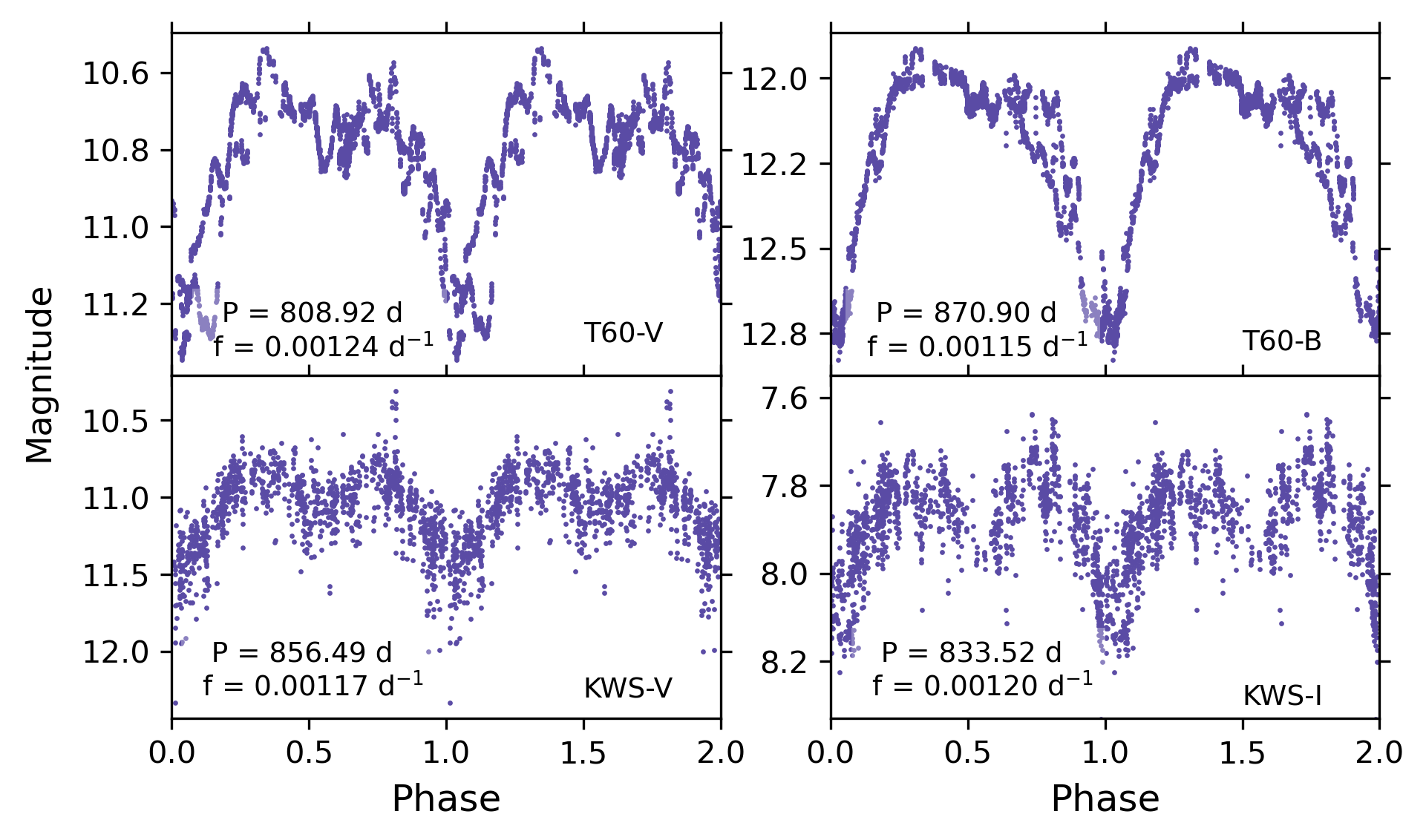}
\caption{Phase-folded light curves of \cicyg, computed using frequencies near the orbital period derived from the periodogram, showing pronounced secondary minima.}
\label{fig:cicyg_second_min}
\end{figure}

\textit{KWS data.}
In the frequency analysis of the KWS-\textit{$I_c$} and \textit{V} bands, the strongest peaks occur at $f = 0.00240 $\rday (416.76~d) and $f = 0.00234 $\rday (428.24~d), representing resonances of the orbital period. The fundamental orbital signal appears as the third-highest peak in the periodogram. When phased with the orbital period, the KWS light curves exhibit a clear secondary minimum, which is likewise evident in the T60 \textit{V}- and \textit{B}-band data Fig.~\ref{fig:cicyg_second_min}. The orbital periods inferred from individual filters show slight variations, primarily due to discrepancies in the positions of these minima.
Another prominent peak appears at $f = 0.00518 $\rday (192.95~d) in the KWS–\textit{Ic} data and $f = 0.00514 $\rday (194.66~d) in the \textit{V}-band. This periodicity is consistent with expectations for pulsational variability of the red giant component.

\begin{figure}
\centering
\includegraphics[width=\linewidth]{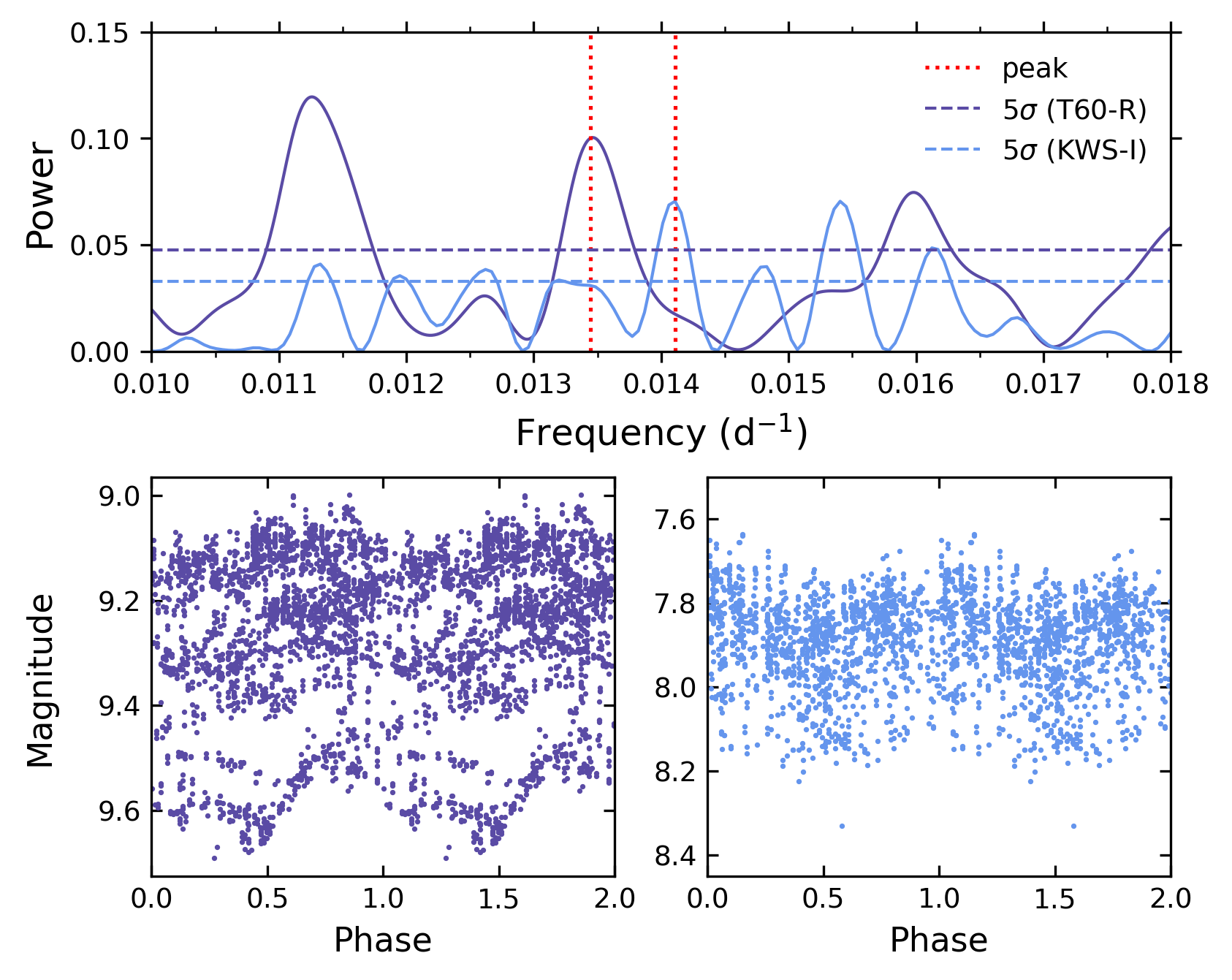}
\caption{Upper panel: Power spectrum zoomed in the range of $0.010-0.018 $\rday for near 70~d variations for T60-\textit{R} and KWS-\textit{$I_c$}. Lower Panel: Folded-phase diagram of $f = 0.01344 $\rday and $f = 0.01411 $\rday showed red dotted lines on the upper panel, left to right, respectively.}
\label{fig:cicyg_70per}
\end{figure}

The “intermediate–amplitude’’ variability of \cicyg is apparent in nearly all observatories and across all photometric bands, though it is most clearly expressed in the \textit{V}, \textit{R}, and \textit{$I_c$} filters (see Fig.~\ref{fig:cicyg_70per}). In the KWS data, the corresponding signals appear at $f = 0.01413 $\rday (70.78~d) in \textit{V} and $f = 0.01411 $\rday (70.88~d) in \textit{$I_c$}. In the \asas data, a comparable frequency of $f = 0.01311 $\rday (76.26~d) is detected.

\textit{TESS data.} The TESS analysis of \cicyg is conducted using the combined light curves from four sectors. Compared to the high variability exhibited by \axper, the light curve of \cicyg is noticeably smoother and flatter. The data display sinusoidal variations with a characteristic timescale of approximately 40~days, which exceeds the duration of the individual sector observations. To isolate shorter-period signals, we remove trends longer than 1~day by detrending the light curve using the flatten routine implemented in the lightkurve package. This method applies Savitzky–Golay filtering with polynomial fits over prescribed windows, effectively normalizing the data and enhancing the visibility of short-timescale modulations. 
The Lomb–Scargle periodogram computed from the resulting detrended light curve reveals a strong and dominant signal within the 0.01–0.1~d range. The most prominent period, 0.046229(6)~d (66.57~min), along with its harmonics, appears across this frequency interval. The corresponding periodogram peaks and the phase-folded light curve at this period are presented in Fig.~\ref{fig:cicyg_tess}. The light curve folded on 66.57~min period displays a clear sinusoidal modulation.

\subsection{Z~And}
\begin{figure*}
    \centering
    \includegraphics[width=0.99\linewidth]{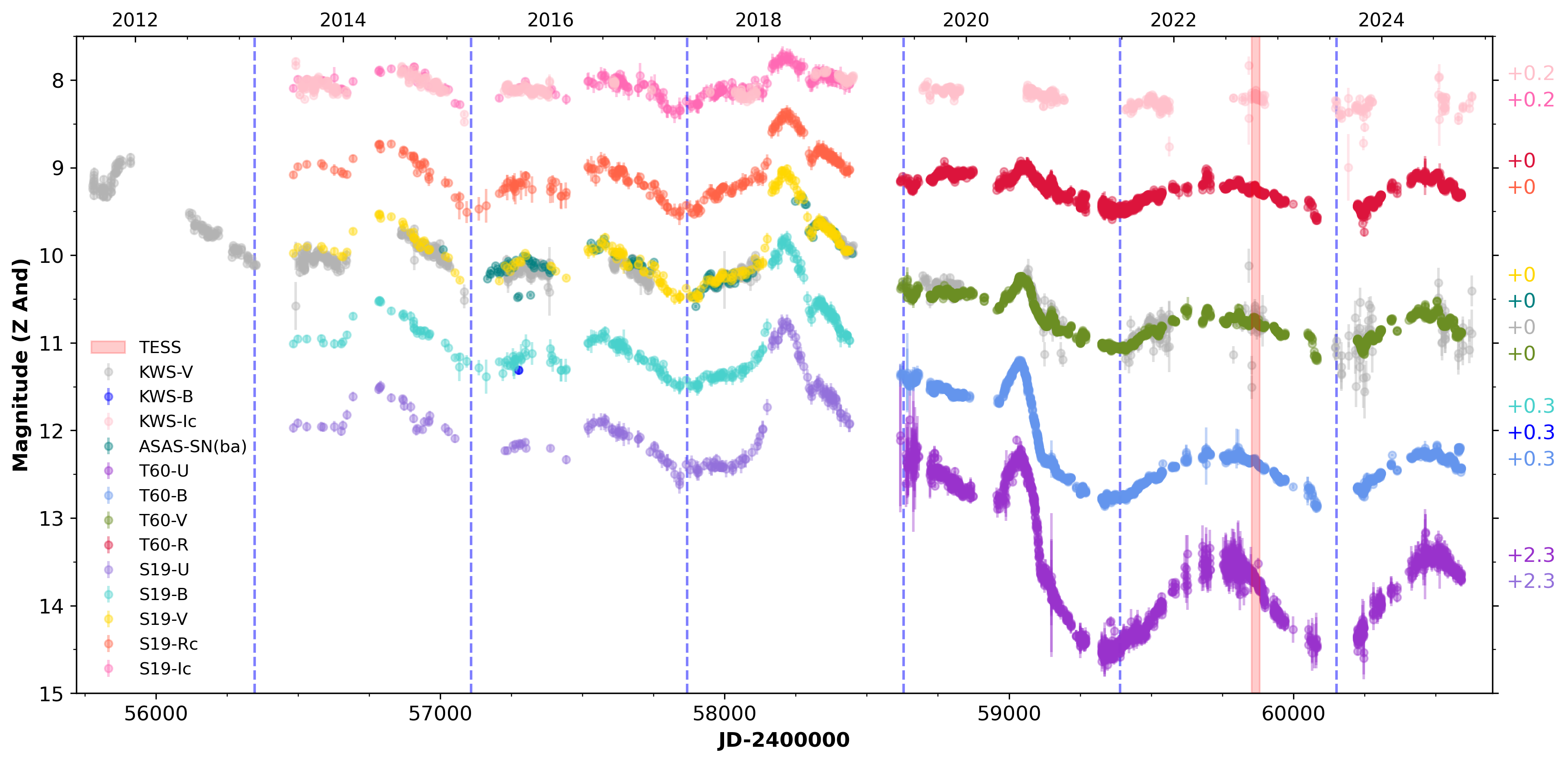}
    \caption{The light variations of all-bands data from \asas, KWS, S19 and T60 for \zand. The red vertical bands span the TESS observation Sector(s), and the blue dashed lines correspond to the photometric eclipsing minima obtained from $\rm JD_{T60\&Fekel}=2450260.2+760.85 \times E$ (see Sec.~\ref{sec:zand_freq}). The offset values with the same color codes are on the outside of the graph.}
    \label{fig:zand_all}
\end{figure*}

\subsubsection{Long-term light curve }
The \textit{V}-band photometric coverage of \zand spans roughly 13.2~yr. Fig.~\ref{fig:zand_all} presents the \textit{V}-band data obtained from KWS, \asas, and T60, with the temporal window of the TESS observations highlighted by the red band. The KWS monitoring begins at a relatively bright epoch of the system ($\rm JD=2455777.2992$), when \zand is observed $V\approx9$~mag.
In the photometric study of \citet{Skopal12}, the onset of this outburst was captured, after 
which their observations concluded. The dataset presented in this work, together with the data of S19,
therefore extends the photometric monitoring beyond the time span covered in that earlier study.

Over the 13~yr monitoring interval, the system exhibits a gradual decline of about one magnitude across two stages. A noticeable outburst occurs at $\rm JD = 2455909.9269$, during which the brightness rises to $V\approx8.8$~mag. This event subsequently fades to $V\approx10.1$~mag over roughly 440~days.
Between $2456351.9015 < \rm JD < 2458107.878$, the system maintains a nearly steady brightness around $V \approx10$~mag, with no major variability. A low-amplitude brightening appears to precede $\rm JD \approx 2456860$, where the light curve shows the faint trailing signature of a decaying outburst.

A second major outburst is detected at $\rm JD \approx 2458315.2$, observed first by \asas and subsequently by KWS, during which the brightness increases from 10.2 to 9.6~mag. A third major outburst occurs $\rm JD \approx 2459066$, reaching a peak brightness of \simi10.2~mag. As the system fades from this event, an additional low-amplitude brightening is observed, after which the system settles into a quiescent state.

\begin{figure}
    \centering
    \includegraphics[width=\linewidth]{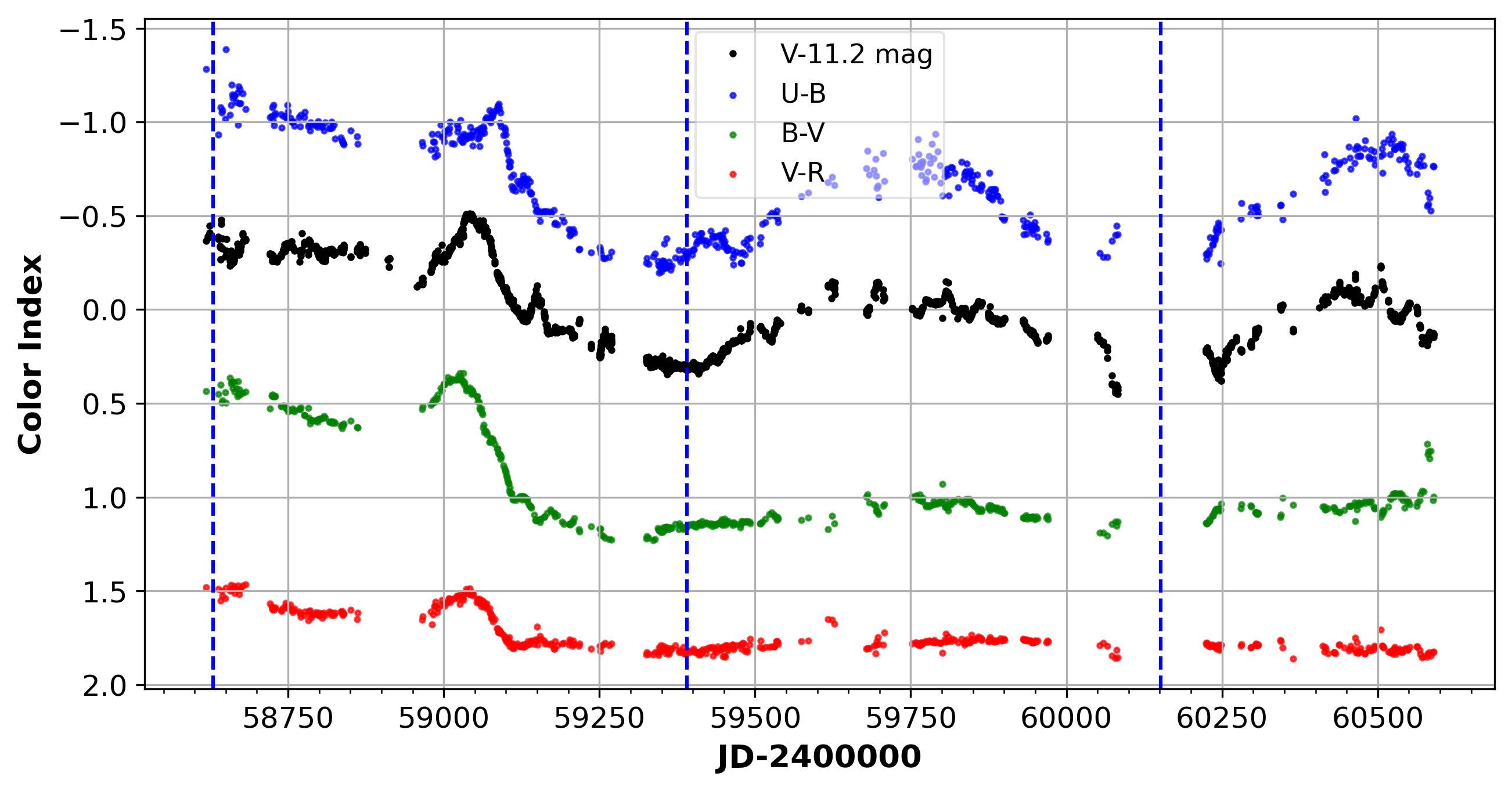}
    \caption{TUG T60 $U-B$, $B-V$ and $V-R$ colors for \zand. T60 \textit{V}-band is also shown by offset by -11.2~mag for easy comparisons with the colors.}
    \label{fig:zand_color}
\end{figure}
\textit{Color-time diagram.}
The mean out-of-eclipse color index decreases over time. Within the interval $2458000 < \rm JD < 2459500$, two consecutive large-amplitude outbursts and one smaller event are detected. Outbursts of this type are characteristic of \zand during its active phases \citep{Merc19}. Throughout this period, the gradual decline in high-energy photon production is clearly visible in Fig.~\ref{fig:zand_color}.

Examining the \textit{U–B} color around the outburst peaking at $\rm JD \sim 2459066$ reveals that the event likely proceeds in two stages. As the outburst rapidly declines, a low-amplitude brightening (originating from the hot component) is observed $\rm JD \sim 2459115$. Prior to the outburst, \textit{U–B} and \textit{B–V} become noticeably bluer. Subsequently, the contribution from ionized gas diminishes, and the flux at shorter wavelengths drops rapidly.

Fig.~\ref{fig:zand_color} marks the phases of the hot-component eclipse with blue dashed lines. Following the eclipse centered $\rm JD \sim 2459390$, the \textit{U–B} color become a bluer lasting roughly 100~d.  The fact that this enhancement appears in \textit{U–B} prior to any noticeable brightening in the \textit{V}-band indicates that the increased energy production in the hot component first manifests through strengthened nebular emission. This rise in ionizing flux most likely originates from localized activity within the inner accretion disk. The near constancy of the other color indexes suggests that the effect is spatially confined. Consequently, the event is consistent with a small-scale outburst driven by a local disk instability.

\subsubsection{TESS observation}
\begin{figure}
\centering
\includegraphics[width=\linewidth]{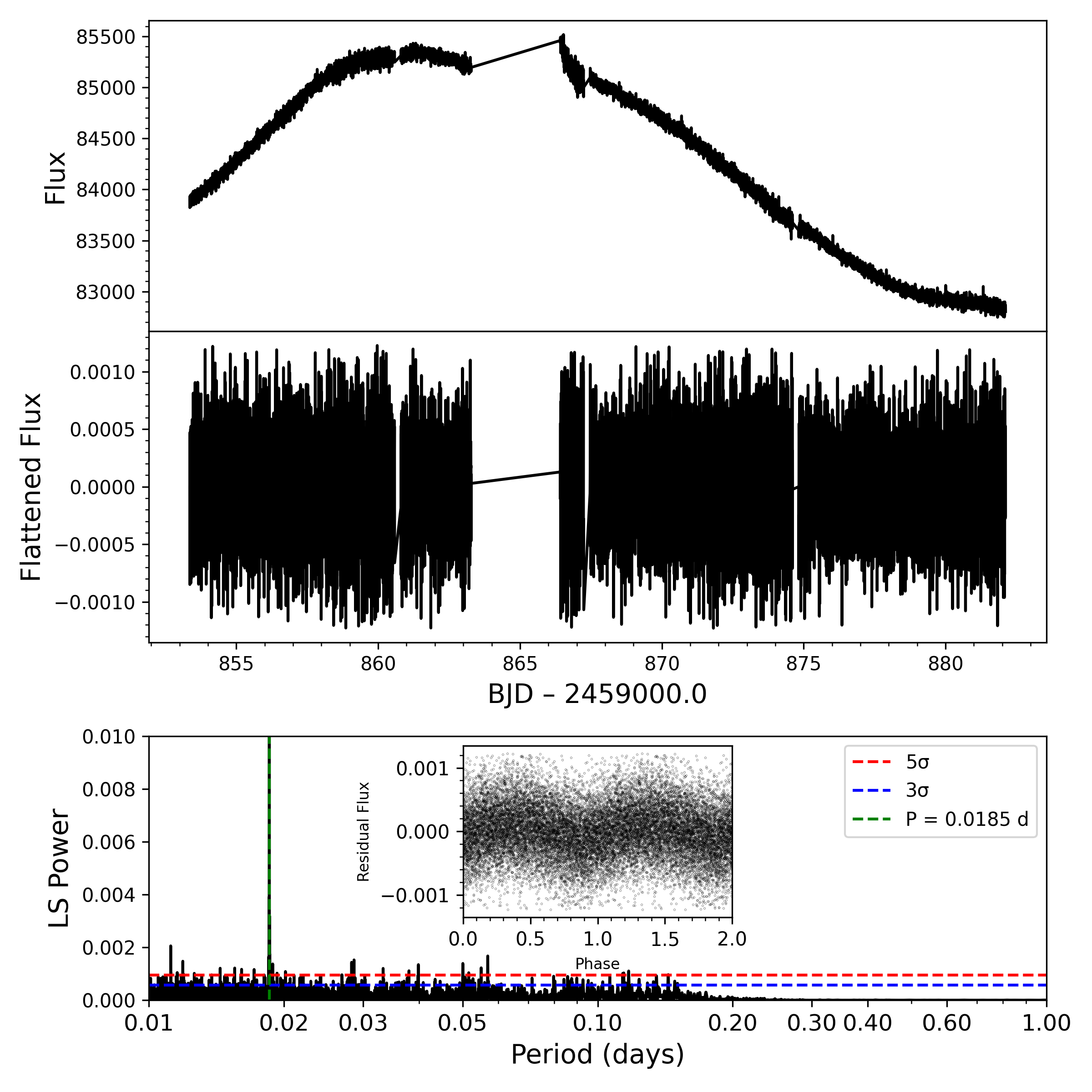}
\caption{TESS photometry and Lomb-Scargle periodogram of \zand. From top to bottom, the background-corrected TESS light curve, the flattened light curve (Savitzky-Golay filter), and the periodogram of the flattened light curve are provided in order. The strongest peak reaches \simi0.3 in amplitude at $P = 0.0185\,\mathrm{d}$ ($f = 54.05\,\mathrm{d^{-1}}$), but the y-axis is limited to 0.01 for clarity.
}
\label{fig:zand_tess}
\end{figure}

As in the case of \axper, only a single TESS dataset is available for \zand. The observations cover the interval $2459853< \rm BJD<2459882$ and correspond to the portion of the long-term light curve in which the system brightens toward a pronounced peak, followed by a rapid decline as the peak sharply subsides. In agreement with this overall behavior, the light curve of Sector~58 exhibits a trend suggesting that the observations capture a segment of a longer-period sinusoidal modulation, as illustrated in Fig.~\ref{fig:zand_tess}.

To extract the short-timescale variability and to enhance the signal amplitudes in the periodogram analyses, we detrend the full Sector~58 light curve using the flatten task defined in lightkurve. The light curve before and after detrending is shown in the top and middle panels of Fig.~\ref{fig:zand_tess}. The resulting residuals were then analyzed with periodograms to search for periodicities in the 0.01–1~d range. The corresponding power spectrum is presented in the bottom panel of Fig.~\ref{fig:zand_tess}.

\begin{figure}
\centering
\includegraphics[width=\linewidth]{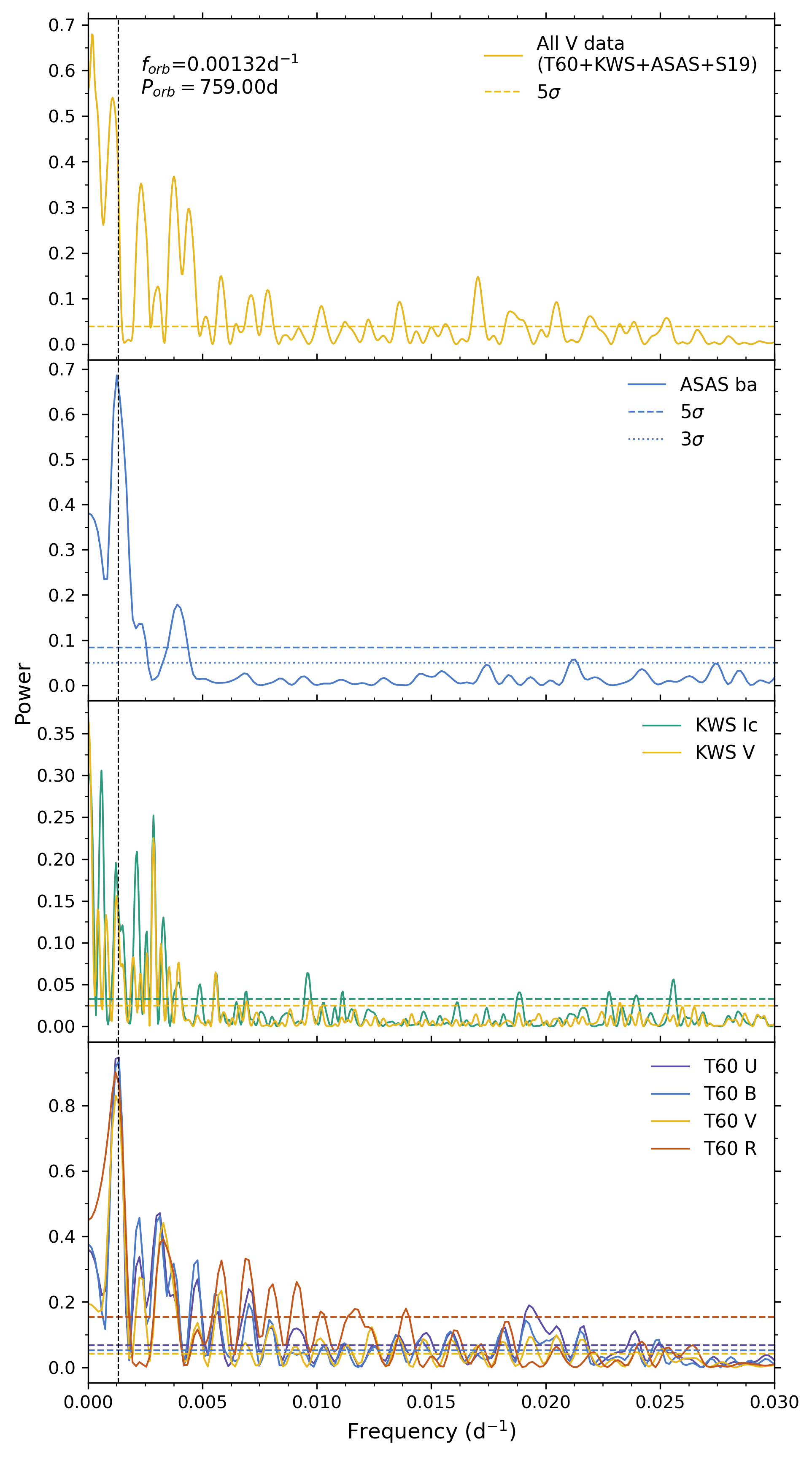}
\caption{Lomb-Scargle Periodogram of \zand for all observations. 
The vertical dashed line denotes the orbital frequency ($f_{\rm orb}$) based on the spectroscopic period of \citet{Fekel00_II}. The horizontal dashed and dotted lines according to the color codes indicate $5\sigma$ and $3\sigma$ significances of corresponding observations, respectively. 
}
\label{fig:zand_freq}
\end{figure}
\subsubsection{Frequency analysis} \label{sec:zand_freq}
A frequency analysis of the 13~yr photometric data of \zand yields a dominant signal \simi800~d, which is larger than the expected orbital period. This offset arises from the system’s active-phase behavior, which introduces irregularities into the long-term light variations. To obtain a more reliable estimate of the orbital modulation, the frequency analysis was restricted to observations obtained after the system entered a quiescent state ($\rm JD > 2459250$). Across all filters, a consistent and well-defined frequency of $f = 0.00131 $\rday (760.85~d) was identified.

Fig.~\ref{fig:zand_freq} presents the resulting power spectrum for the quiescent interval, together with the phase-folded \textit{U}-band light curve at the dominant frequency. The spectroscopic reference epoch $T_0 = 2450260.2$ from \citet{Fekel00_I} was adopted for phasing. The primary-minimum markers in Fig.\ref{fig:zand_color} correspond to the ephemeris $\rm JD_{T60\&Fekel}=2450260.2+760.85 \times E$.

\begin{figure}
\centering
\includegraphics[width=\linewidth]{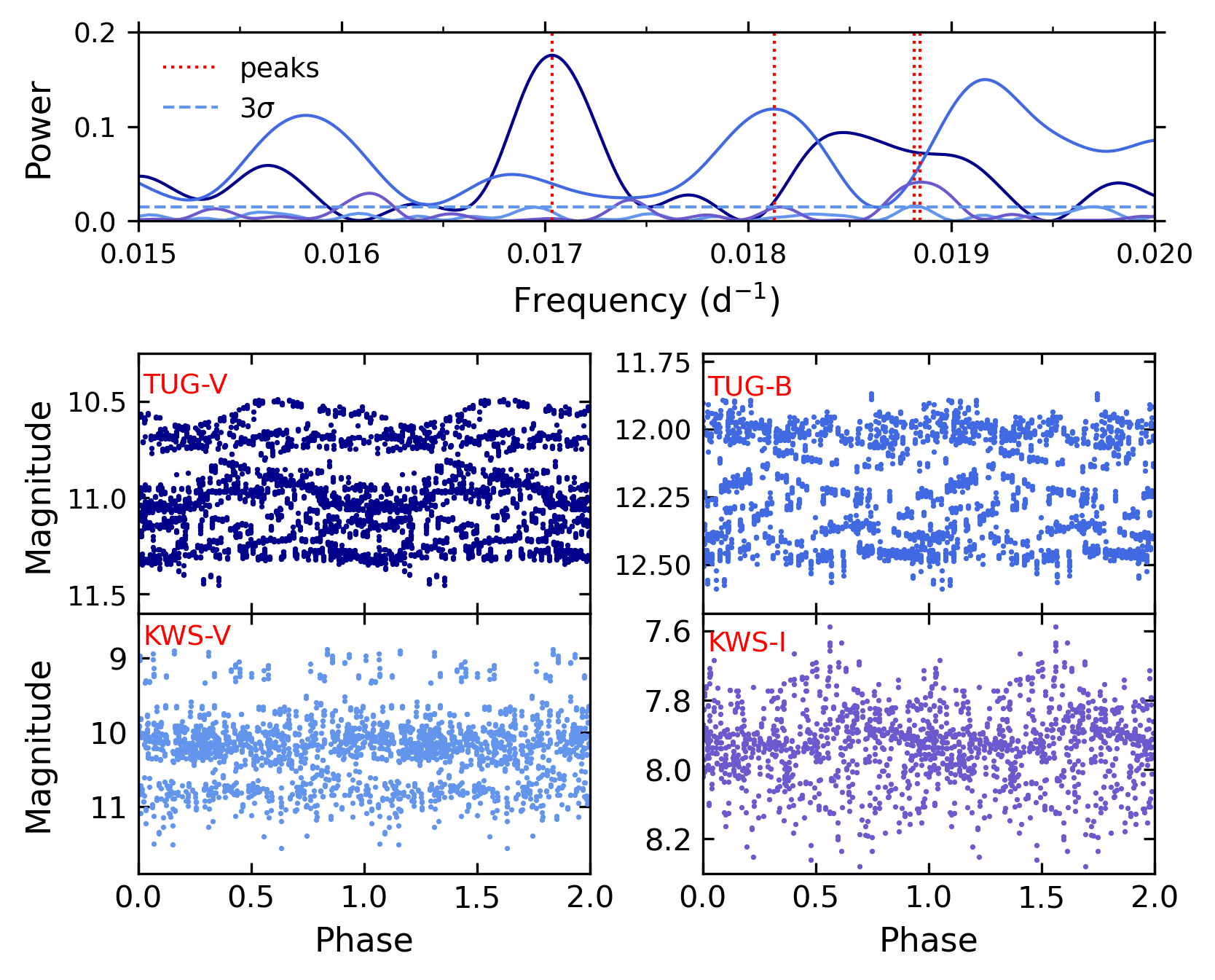}
\caption{Top panel: Power spectrum zoomed in the range of 0.015 -
0.020\rday for around 55~d variations for \zand. 
Lower Panel: The light curves folded by related peak frequencies (see text for details). }
\label{fig:zand_p55}
\end{figure}
\citet{Sekeras19} noted the presence of a \simi58~d modulation in the AAVSO light curve of \zand. Here, we present the frequency analysis of the present data and provide further discussion of this signal. In our analysis of the full T60 data, we detect a clear peak at $f = 0.01704$\rday (58.69~d) in the \textit{V}-band and $f = 0.01703 $\rday (58.71~d) in \textit{R}-band. When the analysis is restricted to the quiescent interval, the characteristic timescale decreases slightly, yielding a period of \simi55~d; the \textit{B}-band data in this interval give $f = 0.01813 $\rday (55.15~d).
Fig.~\ref{fig:zand_p55} shows the corresponding power spectra and phase-folded light curves derived from both the full T60 interval and the quiescent-phase subset. The \asas data independently confirms this modulation, giving $f = 0.01744 $\rday (57.34~d).
A quasi-periodic modulation with a characteristic timescale of \simi58~days becomes prominent across all filters and color indices during the quiescent phase.

Although 55–60~d modulation is less pronounced in the KWS-\textit{V} and \textit{$I_c$} datasets, the dominant peaks occur at $f = 0.01885 $\rday (53.06~d) and $f = 0.01882 $\rday (53.15~d), respectively (see bottom panel of Fig.~\ref{fig:zand_p55}). The longer and more irregular sampling of the KWS observations causes the apparent period to change slightly. Moreover, because \zand is in an active state during roughly 70\% of the KWS coverage, the modulation is significantly diluted by outburst-related variability.

The phase-folded KWS-\textit{V} light curve shows that the system occupies three distinct brightness levels. This multi-level behavior corresponds to the step-like transition from the active to the quiescent state seen in the long-term light curve (Fig.~\ref{fig:zand_all}). In the inner graph of the left-bottom panel of Fig.~\ref{fig:zand_p55}, \simi53~d modulation is more clearly visible during the quiescent interval, when the brightness remains within the 11–10.5~mag range.

Taken together, both the active- and quiescent-state data reveal a semi-periodic modulation. Its enhanced visibility at lower energies and its behavior in the color–time diagrams strongly suggest that this variability arises from non-radial, semi-periodic pulsations of the red giant component.

\textit{TESS data.} 
\zand exhibits a smooth and relatively stable light curve in TESS observations, similar to that of \cicyg. The overall trend indicates that the observations capture a portion of a sector-long, sinusoid-like modulation. After detrending the data, the Lomb–Scargle periodogram reveals a frequency of 0.01854(2)~d (26.7~min), detected within the 0.01–0.1~d interval and accompanied by several harmonics. The light curve folded on this period displays a clear sinusoidal modulation, as presented in Fig.~\ref{fig:zand_tess}.

\begin{table*}
\centering
\caption{Summary of long-, intermediate-, and short-term variability in three symbiotic stars. The Interpretation column lists the possible physical origins corresponding to the long, intermediate, and short modulations, respectively. Here, EM: eclipse-modulated variability; QPO: quasi-periodic oscillation; WD: white dwarf. }
\small
    \begin{tabular}{ccccccc}
    \hline
    \hline
    \bf Name & \bf  Long  & \bf Intermediate & \bf Short & \bf Interpretation \\
             &            & \bf  (30-200~d)  &   &  \\ 
    \hline
    \axper & \simi640~d & 75~d \& 37.7 & \simi0.95~d   & EM; Non-radial puls. \& its harmonic; QPO \\
    \cicyg & \simi855~d & \simi74~d    & \simi66.6~min & EM; Disk+red giant puls.; Possible WD Spin \\
    \zand  & \simi759~d & \simi58.7~d  & 26.7~min      & EM; Accretion-driven variability; WD spin\\
    \hline
    \end{tabular}
    \label{tab:results}
\end{table*}
\section{Discussion}  \label{sec:discussion}
This study presents a comprehensive 13~yr photometric analysis of the symbiotic systems \axper, \cicyg, and \zand, combining long-term TUG~T60 multicolor observations with additional public data sets. T60 observations, which provide the most continuous and highest-quality data within this study, enable the identification of both large-amplitude outbursts and subtle low-amplitude variations occurring on a range of timescales. Given that \sys contain astrophysical components spanning a wide temperature range, including a cool giant, a hot compact object, and an extended ionized nebula, multicolor photometry is essential for diagnosing the physical origin of observed brightness variations. Below, we summarize the physical interpretations of variability for each system and place them in context with earlier results in the literature.

\axper has been in an active stage since mid-2007, consistent with the behavior reported by \citet{Skopal11}. Their observations ended in 2011, whereas the data analyzed here extend from mid-2011 to the end of 2023, effectively continuing the monitoring of the same active phase. Throughout the entire \simi13~yr timespan, the system remained active. 
During the active phase, \axper exhibits narrow and deep minima. However, starting around JD\simi2459000, both the depth and width of minima begin to change, showing shallower and broader profiles. The presence of low-amplitude wave-like fluctuations during eclipses clearly disrupts the originally sharp eclipse shape. In quiescence, \axper is known to exhibit broad minima due to the dense, optically thick ionized wind of the red giant, modulated by orbital motion \citep{Skopal01}. During quiescence, the accretion disk gradually re-forms, leading to the transformation of V-shaped minima into broad structures.
\citet{Skopal11} argued that the 2007–2010 active phase was driven by a significant brightening of the hot component. They modeled the changing eclipse shapes by suggesting that the white dwarf wind accumulates preferentially near the equatorial plane, creating a disk-like structure. This structure reprocesses radiation from the hot component and contributes significantly to the optical brightness. Our T60 multicolor data confirm that \axper remains notably blue, with average colors $U-B \approx 0.75$, $B-V \approx 1.5$, and $V-R \approx -0.2$, consistent with hot-component-dominated emission during this continued active stage.
Orbital period determinations vary across data sets and epochs. While the quiescent-state orbital period is known to be \simi680~d, our combined multi-observatory analysis yields \simi635~d, and the T60 data alone give \simi648.5~d. 
Systems such as \axper are known to exhibit apparent changes in the orbital period, manifested as displacements in the timings of photometric minima caused by changes in the spatial distribution of the dominant optical emission, particularly within the ionized nebula, rather than by true orbital evolution \citep{Skopal98}. In active systems, additional structural complexity (e.g., circumstellar material) may further affect the observed minima. Although differences of several days may appear large in the time domain, they correspond to relatively small offsets in frequency space (\simi$10^{-4} $\rday) and are typical of nebula-embedded long-period symbiotic binaries.
Thus, the main goal here is not to refine the orbital period but to understand the physical mechanisms responsible for the apparent changes in minimum locations.
A striking result from the T60 data is the detection of stable wave-like modulations at 75.46~d and its harmonic 37.73~d. These resonant periods are best explained by non-radial, semi-regular pulsations of the red giant, consistent with expected pulsation ranges of 40–200~d for S-type symbiotics \citep{Merc25}. Such multimode pulsations, possibly modulated by disk instabilities or orbit–pulsation interactions \citep{Leibowitz08,Hinkle19}, also reproduce the color behaviors observed here: brightness increases while colors redden (or vice versa), indicating cool-component variability. All evidence points toward strong, non-radial pulsations of the giant as the primary source of these modulations.

The long-term variability of \cicyg shows a prominent quasi-periodic modulation around \simi70–74~d, with sinusoidal morphology and noticeable amplitude and phase drifts. These low-amplitude variations are strong enough to obscure the primary eclipse minima, making the determination of precise eclipse timings difficult. Despite the similarity to red giant pulsations, \cicyg shows phase-reversed behavior in some colors, implying that more than one physical process may be operating.
Earlier observations by \citet{Kenyon82} reported low-amplitude (0.1–0.3~mag) modulations during eclipse minima, demonstrating that this variability cannot be solely attributed to pulsations or shell burning. Spectroscopic data further suggest contributions from ionized gas structures surrounding the system. If the accretion disk is geometrically larger than the giant at orbital conjunction, the far rim of the disk may remain visible even during eclipse, allowing disk-intrinsic variability to appear as out-of-eclipse-like brightness modulation. For this effect to be strong, the disk must contribute significantly to the UV/optical brightness, consistent with the presence of a hot, optically thin disk during quiescence \citep{Mikolajewska96}.
Given the system’s moderately eccentric orbit \citep[$ e =0.15 \pm 0.08, i =73^\circ \pm 6^\circ $;][]{Kenyon91}, the disk is unlikely to be homogeneous. Non-uniform mass loading from the giant’s wind could naturally produce density waves propagating through the disk, which in turn modulate the emission from the ionized nebula. These considerations support a hybrid interpretation: the \simi74~d variability likely arises from a combination of red giant pulsations and structural/ionization changes within the disk–nebula environment. T60 data show a gradual decline in the modulation amplitude after the second eclipse, consistent with the idea that activity-triggered structural changes from past outbursts may be relaxing. Continued monitoring (especially with spectroscopy)is required to disentangle these contributions.

Across the \simi13~yr dataset, \zand exhibits behavior characteristic of an active system, consistent with long-term activity reported since 2000.7 \citep{Skopal12}. After the brightening around $\rm JD \sim 2459066$, the system began to show broad, orbitally modulated variations similar to quiescent stages, with magnitudes ranging between $V = 10.5-11$. However, as emphasized by \citet{Sekeras19}, \zand can mimic quiescent-like morphology even during active states, making continued observations essential.
\zand is known to display variability from minutes to tens of days. In our T60 analysis, a periodicity near 58.69~d is detected, consistent with the \simi58~d modulation noted in AAVSO \textit{V}-band observations by \citet{Sekeras19}, although they did not explore the origin of this behavior. Our frequency analysis confirms the stability of this timescale, but the physical origin remains unclear. It may relate to quasi-periodic mass-transfer fluctuations, thermal–viscous instabilities, or pulsation-modulated wind variability.

Although not the primary focus of this study, we also present the short-period variations ($<$1~d) identified in the TESS data, which provide a broader context for interpreting the intermediate-timescale variability of the systems.
Detecting short-term variability in \sys requires intensive, multi-color, and preferably continuous overnight observations. Measuring these rapid changes coherently across different energy ranges is essential, as they may be linked to the rotation of the white dwarf. From this perspective, one of the best-characterized systems with a confirmed white dwarf spin period is FN~Sgr, where a period of 11.3~min has been reported \citep{magdolen23}. However, the long orbital periods and strong intrinsic variability typical of symbiotic binaries make the detection of coherent spin signals particularly challenging.

We note that the 26.7~min modulation detected in TESS data of \zand is consistent with the expected spin-related variability of the accreting white dwarf.
This interpretation is strongly supported by the earlier discovery of a persistent \simi28~min optical oscillation by \citet{Sokoloski99}, detected on all observing nights over nearly one year. The close agreement between the two periods suggests a common physical origin, now observed in space-based photometry with improved precision and continuous temporal coverage.
In addition, for \cicyg, we detect a 66.6~min modulation that may plausibly be associated with white dwarf spin. Nonetheless, periodicity of this kind could also arise from plasma motions in the innermost accretion regions, and distinguishing between these scenarios remains nontrivial. For this reason, we strongly recommend dedicated nightly follow-up observations to verify the stability and coherence of the detected frequency.
\citet{Merc24} reported several findings based on TESS observations of some symbiotic stars. Although our analysis employs different methods, these results remain consistent with their conclusions.

In \axper, we identify a signal near \simi0.95~d (22.9~h). Although some cycles of the folded light curve show recognizable structures at this period, the absence of a consistently repeating pattern across all cycles suggests that the variability may be quasi-periodic rather than strictly coherent. Comparisons with magnetic systems such as intermediate polars among cataclysmic variables indicate that confirmed white dwarf spin periods are typically much shorter than one day \citep{demartino20}. However, symbiotic systems 
exhibiting comparatively long spin periods do exist; for example, BF~Cyg shows a modulation of \simi1.8~h \citep{formiggini09}, indicating that extended rotation periods are within the observed range for this class of objects.

Taken as a whole, our results show that \axper, \cicyg, and \zand all display photometric variability spanning several distinct timescales. In each system, long-term activity, intermediate-period modulations, and shorter-period signals are present simultaneously, yet remain clearly separated in period space. Although the observational characteristics of these variations have now been identified, their physical origin and possible interconnections are not immediately evident. In the next section, we therefore examine these results in a broader physical context, considering the roles of red-giant pulsations, accretion-related phenomena, and the overall structure of symbiotic binary systems.

\section{Conclusion}
The three systems display a rich hierarchy of variability spanning timescales from minutes to years. While the long-term behavior is governed by orbital modulation and outburst activity, all targets exhibit persistent intermediate-timescale oscillations with periods between 30 and 100 days. These signals appear consistently across independent datasets, persist through both active and quiescent states, and reach amplitudes of up to 0.25~mag. Their presence in all three systems points to a common physical origin, potentially involving large-scale variations in the red-giant wind, magnetically influenced accretion structures, or quasi-periodic density enhancements at the wind–accretor interaction interface.

The observed periods fall outside the range of the fundamental and overtone radial pulsations expected for M giants, yet they are naturally compatible with semi-regular or non-radial pulsation modes of the cool component. Such pulsations may lead to time-dependent variations in the mass-loss rate from the red giant and, consequently, to transient enhancements in the accretion rate onto the white dwarf.
This scenario provides a plausible framework linking intermediate-timescale variability to the triggering of Z~And-type outbursts, although a direct causal connection has not yet been observationally demonstrated.
In \axper and \cicyg, the persistence of 37–75~d signals across multiple studies indicates a long-lived and possibly stable process. In \zand, the \simi58~d modulation, now confirmed independently, demonstrates that such intermediate-timescale variability is a robust and intrinsic component of symbiotic mass transfer.

High-cadence TESS photometry provides compelling evidence for short-period variability in these systems. The 26.7~min modulation in \zand and the 66.6~min signal in \cicyg remain phase-coherent over multiple sectors, supporting interpretations involving white dwarf spin or magnetically gated accretion. 
The weak 0.95~d signal in \axper is confined to quiescent intervals and therefore requires independent verification. Together, these detections show that the hot components of symbiotic binaries can sustain stable or quasi-stable rotational or accretion-driven oscillations that are accessible only through continuous space-based monitoring. 

Taken together, our results support the view that intermediate-timescale photometric variations on the order of tens of days are a recurrent and physically significant component of variability in shell-burning symbiotic binaries.
Future multiwavelength campaigns that combine optical photometry, UV spectroscopy, and hydrodynamical modeling will be essential to determine whether the dominant mechanism arises from pulsation, wind-structure variations, or accretion-related instabilities. Extending the analysis to additional symbiotic systems observed by TESS will further clarify whether these modulations are truly widespread.

\begin{acknowledgments}
The authors thank the anonymous referee for their constructive comments and valuable suggestions, which significantly improved the quality of this manuscript. 
This study was supported by the Scientific and Technological Research Council of Türkiye (TÜBİTAK) under Grant Number 124F073. The authors thank TÜBİTAK for their support.
We thank TÜBİTAK National Observatory (TUG) for partial support in using T60 telescope with project number 19BT60-1495. This article includes a part of the PhD thesis of M.Y.
\end{acknowledgments}





%
\facilities{TUG~T60: 60cm, \asas, KWS, TESS, Star\'a Lesn\'a Observatory, Gaia}

\software{
astropy \citep{2013A&A...558A..33A,2018AJ....156..123A,2022ApJ...935..167A}, AstroImageJ \citep{Collins17},
LombScargle \citep{Scargle82,Vanderplas18}, 
lightkurve \citep{lightkurve18}, 
TESSCut \citep{Brasseur19}
}





\bibliography{syst}{}

@ARTICLE{2022ApJ...935..167A,
       author = {{Astropy Collaboration} and {Price-Whelan}, Adrian M. and {Lim}, Pey Lian and {Earl}, Nicholas and {Starkman}, Nathaniel and {Bradley}, Larry and {Shupe}, David L. and {Patil}, Aarya A. and {Corrales}, Lia and {Brasseur}, C.~E. and {N{\"o}the}, Maximilian and {Donath}, Axel and {Tollerud}, Erik and {Morris}, Brett M. and {Ginsburg}, Adam and {Vaher}, Eero and {Weaver}, Benjamin A. and {Tocknell}, James and {Jamieson}, William and {van Kerkwijk}, Marten H. and {Robitaille}, Thomas P. and {Merry}, Bruce and {Bachetti}, Matteo and {G{\"u}nther}, H. Moritz and {Aldcroft}, Thomas L. and {Alvarado-Montes}, Jaime A. and {Archibald}, Anne M. and {B{\'o}di}, Attila and {Bapat}, Shreyas and {Barentsen}, Geert and {Baz{\'a}n}, Juanjo and {Biswas}, Manish and {Boquien}, M{\'e}d{\'e}ric and {Burke}, D.~J. and {Cara}, Daria and {Cara}, Mihai and {Conroy}, Kyle E. and {Conseil}, Simon and {Craig}, Matthew W. and {Cross}, Robert M. and {Cruz}, Kelle L. and {D'Eugenio}, Francesco and {Dencheva}, Nadia and {Devillepoix}, Hadrien A.~R. and {Dietrich}, J{\"o}rg P. and {Eigenbrot}, Arthur Davis and {Erben}, Thomas and {Ferreira}, Leonardo and {Foreman-Mackey}, Daniel and {Fox}, Ryan and {Freij}, Nabil and {Garg}, Suyog and {Geda}, Robel and {Glattly}, Lauren and {Gondhalekar}, Yash and {Gordon}, Karl D. and {Grant}, David and {Greenfield}, Perry and {Groener}, Austen M. and {Guest}, Steve and {Gurovich}, Sebastian and {Handberg}, Rasmus and {Hart}, Akeem and {Hatfield-Dodds}, Zac and {Homeier}, Derek and {Hosseinzadeh}, Griffin and {Jenness}, Tim and {Jones}, Craig K. and {Joseph}, Prajwel and {Kalmbach}, J. Bryce and {Karamehmetoglu}, Emir and {Ka{\l}uszy{\'n}ski}, Miko{\l}aj and {Kelley}, Michael S.~P. and {Kern}, Nicholas and {Kerzendorf}, Wolfgang E. and {Koch}, Eric W. and {Kulumani}, Shankar and {Lee}, Antony and {Ly}, Chun and {Ma}, Zhiyuan and {MacBride}, Conor and {Maljaars}, Jakob M. and {Muna}, Demitri and {Murphy}, N.~A. and {Norman}, Henrik and {O'Steen}, Richard and {Oman}, Kyle A. and {Pacifici}, Camilla and {Pascual}, Sergio and {Pascual-Granado}, J. and {Patil}, Rohit R. and {Perren}, Gabriel I. and {Pickering}, Timothy E. and {Rastogi}, Tanuj and {Roulston}, Benjamin R. and {Ryan}, Daniel F. and {Rykoff}, Eli S. and {Sabater}, Jose and {Sakurikar}, Parikshit and {Salgado}, Jes{\'u}s and {Sanghi}, Aniket and {Saunders}, Nicholas and {Savchenko}, Volodymyr and {Schwardt}, Ludwig and {Seifert-Eckert}, Michael and {Shih}, Albert Y. and {Jain}, Anany Shrey and {Shukla}, Gyanendra and {Sick}, Jonathan and {Simpson}, Chris and {Singanamalla}, Sudheesh and {Singer}, Leo P. and {Singhal}, Jaladh and {Sinha}, Manodeep and {Sip{\H{o}}cz}, Brigitta M. and {Spitler}, Lee R. and {Stansby}, David and {Streicher}, Ole and {{\v{S}}umak}, Jani and {Swinbank}, John D. and {Taranu}, Dan S. and {Tewary}, Nikita and {Tremblay}, Grant R. and {de Val-Borro}, Miguel and {Van Kooten}, Samuel J. and {Vasovi{\'c}}, Zlatan and {Verma}, Shresth and {de Miranda Cardoso}, Jos{\'e} Vin{\'\i}cius and {Williams}, Peter K.~G. and {Wilson}, Tom J. and {Winkel}, Benjamin and {Wood-Vasey}, W.~M. and {Xue}, Rui and {Yoachim}, Peter and {Zhang}, Chen and {Zonca}, Andrea and {Astropy Project Contributors}},
        title = "{The Astropy Project: Sustaining and Growing a Community-oriented Open-source Project and the Latest Major Release (v5.0) of the Core Package}",
      journal = {\apj},
         year = 2022,
        month = aug,
       volume = {935},
       number = {2},
          eid = {167},
        pages = {167},
          doi = {10.3847/1538-4357/ac7c74},
archivePrefix = {arXiv},
       eprint = {2206.14220},
 primaryClass = {astro-ph.IM},
       adsurl = {https://ui.adsabs.harvard.edu/abs/2022ApJ...935..167A}
}

@ARTICLE{2018AJ....156..123A,
       author = {{Astropy Collaboration} and {Price-Whelan}, A.~M. and {Sip{\H{o}}cz}, B.~M. and {G{\"u}nther}, H.~M. and {Lim}, P.~L. and {Crawford}, S.~M. and {Conseil}, S. and {Shupe}, D.~L. and {Craig}, M.~W. and {Dencheva}, N. and {Ginsburg}, A. and {VanderPlas}, J.~T. and {Bradley}, L.~D. and {P{\'e}rez-Su{\'a}rez}, D. and {de Val-Borro}, M. and {Aldcroft}, T.~L. and {Cruz}, K.~L. and {Robitaille}, T.~P. and {Tollerud}, E.~J. and {Ardelean}, C. and {Babej}, T. and {Bach}, Y.~P. and {Bachetti}, M. and {Bakanov}, A.~V. and {Bamford}, S.~P. and {Barentsen}, G. and {Barmby}, P. and {Baumbach}, A. and {Berry}, K.~L. and {Biscani}, F. and {Boquien}, M. and {Bostroem}, K.~A. and {Bouma}, L.~G. and {Brammer}, G.~B. and {Bray}, E.~M. and {Breytenbach}, H. and {Buddelmeijer}, H. and {Burke}, D.~J. and {Calderone}, G. and {Cano Rodr{\'\i}guez}, J.~L. and {Cara}, M. and {Cardoso}, J.~V.~M. and {Cheedella}, S. and {Copin}, Y. and {Corrales}, L. and {Crichton}, D. and {D'Avella}, D. and {Deil}, C. and {Depagne}, {\'E}. and {Dietrich}, J.~P. and {Donath}, A. and {Droettboom}, M. and {Earl}, N. and {Erben}, T. and {Fabbro}, S. and {Ferreira}, L.~A. and {Finethy}, T. and {Fox}, R.~T. and {Garrison}, L.~H. and {Gibbons}, S.~L.~J. and {Goldstein}, D.~A. and {Gommers}, R. and {Greco}, J.~P. and {Greenfield}, P. and {Groener}, A.~M. and {Grollier}, F. and {Hagen}, A. and {Hirst}, P. and {Homeier}, D. and {Horton}, A.~J. and {Hosseinzadeh}, G. and {Hu}, L. and {Hunkeler}, J.~S. and {Ivezi{\'c}}, {\v{Z}}. and {Jain}, A. and {Jenness}, T. and {Kanarek}, G. and {Kendrew}, S. and {Kern}, N.~S. and {Kerzendorf}, W.~E. and {Khvalko}, A. and {King}, J. and {Kirkby}, D. and {Kulkarni}, A.~M. and {Kumar}, A. and {Lee}, A. and {Lenz}, D. and {Littlefair}, S.~P. and {Ma}, Z. and {Macleod}, D.~M. and {Mastropietro}, M. and {McCully}, C. and {Montagnac}, S. and {Morris}, B.~M. and {Mueller}, M. and {Mumford}, S.~J. and {Muna}, D. and {Murphy}, N.~A. and {Nelson}, S. and {Nguyen}, G.~H. and {Ninan}, J.~P. and {N{\"o}the}, M. and {Ogaz}, S. and {Oh}, S. and {Parejko}, J.~K. and {Parley}, N. and {Pascual}, S. and {Patil}, R. and {Patil}, A.~A. and {Plunkett}, A.~L. and {Prochaska}, J.~X. and {Rastogi}, T. and {Reddy Janga}, V. and {Sabater}, J. and {Sakurikar}, P. and {Seifert}, M. and {Sherbert}, L.~E. and {Sherwood-Taylor}, H. and {Shih}, A.~Y. and {Sick}, J. and {Silbiger}, M.~T. and {Singanamalla}, S. and {Singer}, L.~P. and {Sladen}, P.~H. and {Sooley}, K.~A. and {Sornarajah}, S. and {Streicher}, O. and {Teuben}, P. and {Thomas}, S.~W. and {Tremblay}, G.~R. and {Turner}, J.~E.~H. and {Terr{\'o}n}, V. and {van Kerkwijk}, M.~H. and {de la Vega}, A. and {Watkins}, L.~L. and {Weaver}, B.~A. and {Whitmore}, J.~B. and {Woillez}, J. and {Zabalza}, V. and {Astropy Contributors}},
        title = "{The Astropy Project: Building an Open-science Project and Status of the v2.0 Core Package}",
      journal = {\aj},
         year = 2018,
        month = sep,
       volume = {156},
       number = {3},
          eid = {123},
        pages = {123},
          doi = {10.3847/1538-3881/aabc4f},
archivePrefix = {arXiv},
       eprint = {1801.02634},
 primaryClass = {astro-ph.IM},
       adsurl = {https://ui.adsabs.harvard.edu/abs/2018AJ....156..123A}
}

@ARTICLE{2013A&A...558A..33A,
       author = {{Astropy Collaboration} and {Robitaille}, Thomas P. and
         {Tollerud}, Erik J. and {Greenfield}, Perry and {Droettboom}, Michael and
         {Bray}, Erik and {Aldcroft}, Tom and {Davis}, Matt and
         {Ginsburg}, Adam and {Price-Whelan}, Adrian M. and
         {Kerzendorf}, Wolfgang E. and {Conley}, Alexander and {Crighton}, Neil and
         {Barbary}, Kyle and {Muna}, Demitri and {Ferguson}, Henry and
         {Grollier}, Fr{\'e}d{\'e}ric and {Parikh}, Madhura M. and
         {Nair}, Prasanth H. and {Unther}, Hans M. and {Deil}, Christoph and
         {Woillez}, Julien and {Conseil}, Simon and {Kramer}, Roban and
         {Turner}, James E.~H. and {Singer}, Leo and {Fox}, Ryan and
         {Weaver}, Benjamin A. and {Zabalza}, Victor and {Edwards}, Zachary I. and
         {Azalee Bostroem}, K. and {Burke}, D.~J. and {Casey}, Andrew R. and
         {Crawford}, Steven M. and {Dencheva}, Nadia and {Ely}, Justin and
         {Jenness}, Tim and {Labrie}, Kathleen and {Lim}, Pey Lian and
         {Pierfederici}, Francesco and {Pontzen}, Andrew and {Ptak}, Andy and
         {Refsdal}, Brian and {Servillat}, Mathieu and {Streicher}, Ole},
        title = "{Astropy: A community Python package for astronomy}",
      journal = {\aap},
         year = "2013",
        month = "Oct",
       volume = {558},
          eid = {A33},
        pages = {A33},
          doi = {10.1051/0004-6361/201322068},
archivePrefix = {arXiv},
       eprint = {1307.6212},
 primaryClass = {astro-ph.IM},
       adsurl = {https://ui.adsabs.harvard.edu/abs/2013A&A...558A..33A}
}

@ARTICLE{Hinkle19,
       author = {{Hinkle}, Kenneth H. and {Fekel}, Francis C. and {Joyce}, Richard R. and {Miko{\l}ajewska}, Joanna and {Ga{\l}an}, Cezary and {Lebzelter}, Thomas},
        title = "{Infrared Spectroscopy of Symbiotic Stars. XII. The Neutron Star SyXB System 4U 1700+24 = V934 Herculis}",
      journal = {\apj},
         year = 2019,
        month = feb,
       volume = {872},
       number = {1},
          eid = {43},
        pages = {43},
          doi = {10.3847/1538-4357/aafba5},
archivePrefix = {arXiv},
       eprint = {1812.08811},
 primaryClass = {astro-ph.SR},
       adsurl = {https://ui.adsabs.harvard.edu/abs/2019ApJ...872...43H}
}

@ARTICLE{Belczynski00,
       author = {{Belczy{\'n}ski}, K. and {Miko{\l}ajewska}, J. and {Munari}, U. and {Ivison}, R.~J. and {Friedjung}, M.},
        title = "{A catalogue of symbiotic stars}",
      journal = {\aaps},
         year = 2000,
        month = nov,
       volume = {146},
        pages = {407-435},
          doi = {10.1051/aas:2000280},
archivePrefix = {arXiv},
       eprint = {astro-ph/0005547},
 primaryClass = {astro-ph},
       adsurl = {https://ui.adsabs.harvard.edu/abs/2000A&AS..146..407B}
}

@ARTICLE{Skopal12,
       author = {{Skopal}, A. and {Shugarov}, S. and {Va{\v{n}}ko}, M. and {Dubovsk{\'y}}, P. and {Peneva}, S.~P. and {Semkov}, E. and {Wolf}, M.},
        title = "{Recent photometry of symbiotic stars}",
      journal = {Astronomische Nachrichten},
         year = 2012,
        month = apr,
       volume = {333},
       number = {3},
        pages = {242},
          doi = {10.1002/asna.201111655},
archivePrefix = {arXiv},
       eprint = {1203.4932},
 primaryClass = {astro-ph.SR},
       adsurl = {https://ui.adsabs.harvard.edu/abs/2012AN....333..242S}
}

@ARTICLE{Sokoloski06a,
       author = {{Sokoloski}, J.~L. and {Kenyon}, S.~J. and {Espey}, B.~R. and {Keyes}, Charles D. and {McCandliss}, S.~R. and {Kong}, A.~K.~H. and {Aufdenberg}, J.~P. and {Filippenko}, A.~V. and {Li}, W. and {Brocksopp}, C. and {Kaiser}, Christian R. and {Charles}, P.~A. and {Rupen}, M.~P. and {Stone}, R.~P.~S.},
        title = "{A ``Combination Nova'' Outburst in Z Andromedae: Nuclear Shell Burning Triggered by a Disk Instability}",
      journal = {\apj},
         year = 2006,
        month = jan,
       volume = {636},
       number = {2},
        pages = {1002-1019},
          doi = {10.1086/498206},
archivePrefix = {arXiv},
       eprint = {astro-ph/0509638},
 primaryClass = {astro-ph},
       adsurl = {https://ui.adsabs.harvard.edu/abs/2006ApJ...636.1002S}
}

@ARTICLE{Sokoloski06b,
       author = {{Sokoloski}, J.~L. and {Luna}, G.~J.~M. and {Mukai}, K. and {Kenyon}, Scott J.},
        title = "{An X-ray-emitting blast wave from the recurrent nova RS Ophiuchi}",
      journal = {\nat},
         year = 2006,
        month = jul,
       volume = {442},
       number = {7100},
        pages = {276-278},
          doi = {10.1038/nature04893},
archivePrefix = {arXiv},
       eprint = {astro-ph/0605326},
 primaryClass = {astro-ph},
       adsurl = {https://ui.adsabs.harvard.edu/abs/2006Natur.442..276S}
}

@ARTICLE{Skopal17,
       author = {{Skopal}, A. and {Shugarov}, S. Yu. and {Seker{\'a}{\v{s}}}, M. and {Wolf}, M. and {Tarasova}, T.~N. and {Teyssier}, F. and {Fujii}, M. and {Guarro}, J. and {Garde}, O. and {Graham}, K. and {Lester}, T. and {Bouttard}, V. and {Lemoult}, T. and {Sollecchia}, U. and {Montier}, J. and {Boyd}, D.},
        title = "{New outburst of the symbiotic nova AG Pegasi after 165 yr}",
      journal = {\aap},
         year = 2017,
        month = aug,
       volume = {604},
          eid = {A48},
        pages = {A48},
          doi = {10.1051/0004-6361/201629593},
archivePrefix = {arXiv},
       eprint = {1705.00076},
 primaryClass = {astro-ph.SR},
       adsurl = {https://ui.adsabs.harvard.edu/abs/2017A&A...604A..48S}
}

@ARTICLE{Skopal18,
       author = {{Skopal}, Augustin and {Tarasova}, Taya. N. and {Wolf}, Marek and {Dubovsk{\'y}}, Pavol A. and {Kudzej}, Igor},
        title = "{Repeated Transient Jets from a Warped Disk in the Symbiotic Prototype Z And: A Link to the Long-lasting Active Phase}",
      journal = {\apj},
         year = 2018,
        month = may,
       volume = {858},
       number = {2},
          eid = {120},
        pages = {120},
          doi = {10.3847/1538-4357/aabc11},
archivePrefix = {arXiv},
       eprint = {1805.10908},
 primaryClass = {astro-ph.SR},
       adsurl = {https://ui.adsabs.harvard.edu/abs/2018ApJ...858..120S}
}

@ARTICLE{Skopal20,
       author = {{Skopal}, A. and {Shugarov}, S. Yu. and {Munari}, U. and {Masetti}, N. and {Marchesini}, E. and {Kom{\v{z}}{\'\i}k}, R.~M. and {Kundra}, E. and {Shagatova}, N. and {Tarasova}, T.~N. and {Buil}, C. and {Boussin}, C. and {Shenavrin}, V.~I. and {Hambsch}, F.-J. and {Dallaporta}, S. and {Frigo}, A. and {Garde}, O. and {Zubareva}, A. and {Dubovsk{\'y}}, P.~A. and {Kroll}, P.},
        title = "{The path to Z And-type outbursts: The case of V426 Sagittae (HBHA 1704-05)}",
      journal = {\aap},
         year = 2020,
        month = apr,
       volume = {636},
          eid = {A77},
        pages = {A77},
          doi = {10.1051/0004-6361/201937199},
archivePrefix = {arXiv},
       eprint = {2003.10135},
 primaryClass = {astro-ph.SR},
       adsurl = {https://ui.adsabs.harvard.edu/abs/2020A&A...636A..77S}
}

@ARTICLE{Sokoloski99,
       author = {{Sokoloski}, J.~L. and {Bildsten}, Lars},
        title = "{Discovery of a Magnetic White Dwarf in the Symbiotic Binary Z Andromedae}",
      journal = {\apj},
         year = 1999,
        month = jun,
       volume = {517},
       number = {2},
        pages = {919-924},
          doi = {10.1086/307234},
archivePrefix = {arXiv},
       eprint = {astro-ph/9812294},
 primaryClass = {astro-ph},
       adsurl = {https://ui.adsabs.harvard.edu/abs/1999ApJ...517..919S}
}

@ARTICLE{Formiggini94,
       author = {{Formiggini}, L. and {Leibowitz}, E.~M.},
        title = "{Three periodicities in a 98-year light curve of the symbiotic star Z Andromedae.}",
      journal = {\aap},
         year = 1994,
        month = dec,
       volume = {292},
        pages = {534-542},
       adsurl = {https://ui.adsabs.harvard.edu/abs/1994A&A...292..534F}
}

@BOOK{Kenyon86,
       author = {{Kenyon}, S.~J.},
        title = "{The symbiotic stars}",
         year = 1986,
       publisher = {Cambridge University Press}, 
       ISBN = {9780511586071 (online)},
       doi = {10.1017/CBO9780511586071},
       adsurl = {https://ui.adsabs.harvard.edu/abs/1986syst.book.....K}
}

@ARTICLE{Sekeras19,
       author = {{Seker{\'a}{\v{s}}}, M. and {Skopal}, A. and {Shugarov}, S. and {Shagatova}, N. and {Kundra}, E. and {Kom{\v{z}}{\'\i}k}, R. and {Vra{\v{s}}{\v{t}}{\'a}k}, M. and {Peneva}, S.~P. and {Semkov}, E. and {Stubbings}, R.},
        title = "{Photometry of Symbiotic Stars - XIV}",
      journal = {Contributions of the Astronomical Observatory Skalnate Pleso},
         year = 2019,
        month = apr,
       volume = {49},
       number = {1},
        pages = {19-66},
archivePrefix = {arXiv},
       eprint = {1904.05555},
 primaryClass = {astro-ph.SR},
       adsurl = {https://ui.adsabs.harvard.edu/abs/2019CoSka..49...19S}
}

@ARTICLE{Skopal11,
       author = {{Skopal}, A. and {Tarasova}, T.~N. and {Carikov{\'a}}, Z. and {Castellani}, F. and {Cherini}, G. and {Dallaporta}, S. and {Frigo}, A. and {Marangoni}, C. and {Moretti}, S. and {Munari}, U. and {Righetti}, G.~L. and {Siviero}, A. and {Tomaselli}, S. and {Vagnozzi}, A. and {Valisa}, P.},
        title = "{Formation of a disk structure in the symbiotic binary AX Persei during its 2007-10 precursor-type activity}",
      journal = {\aap},
         year = 2011,
        month = dec,
       volume = {536},
          eid = {A27},
        pages = {A27},
          doi = {10.1051/0004-6361/201116969},
archivePrefix = {arXiv},
       eprint = {1110.2801},
 primaryClass = {astro-ph.SR},
       adsurl = {https://ui.adsabs.harvard.edu/abs/2011A&A...536A..27S}
}

@ARTICLE{Muerset91,
       author = {{M{\"u}rset}, U. and {Nussbaumer}, H. and {Schmid}, H.~M. and {Vogel}, M.},
        title = "{Temperature and luminosity of hot components in symbiotic stars.}",
      journal = {\aap},
         year = 1991,
        month = aug,
       volume = {248},
        pages = {458},
       adsurl = {https://ui.adsabs.harvard.edu/abs/1991A&A...248..458M}
}

@ARTICLE{Kenyon91,
       author = {{Kenyon}, S.~J. and {Oliversen}, N.~A. and {Mikolajewska}, J. and {Mikolajewski}, M. and {Stencel}, R.~E. and {Garcia}, M.~R. and {Anderson}, C.~M.},
        title = "{On the Nature of the Symbiotic Binary CI Cygni}",
      journal = {\aj},
         year = 1991,
        month = feb,
       volume = {101},
        pages = {637},
          doi = {10.1086/115712},
       adsurl = {https://ui.adsabs.harvard.edu/abs/1991AJ....101..637K}
}

@INPROCEEDINGS{Mikolajewska03,
       author = {{Miko{\l}ajewska}, J. and {Kolotilov}, E.~A. and {Shugarov}, S. Yu. and {Tatarnikova}, A.~A. and {Yudin}, B.~F.},
        title = "{Ellipsoidal Variability of Symbiotic Giants}",
    booktitle = {Symbiotic Stars Probing Stellar Evolution},
         year = 2003,
       editor = {{Corradi}, R.~L.~M. and {Mikolajewska}, J. and {Mahoney}, T.~J.},
       series = {Astronomical Society of the Pacific Conference Series},
       volume = {303},
        month = jan,
        pages = {151},
archivePrefix = {arXiv},
       eprint = {astro-ph/0210492},
 primaryClass = {astro-ph},
       adsurl = {https://ui.adsabs.harvard.edu/abs/2003ASPC..303..151M}
}

@ARTICLE{Maehara14,
       author = {{Maehara}, H.},
        title = "{Automated Wide-field Survey for Transient Objects with a Small Telescope}",
      journal = {Journal of Space Science Informatics Japan},
         year = 2014,
        month = mar,
       volume = {3},
        pages = {119-127},
          doi = {},
archivePrefix = {},
       eprint = {},
 primaryClass = {},
       adsurl = {}
}

@ARTICLE{King09,
       author = {{King}, A.~R. and {Pringle}, J.~E.},
        title = "{RS Ophiuchi: thermonuclear explosion or disc instability?}",
      journal = {\mnras},
         year = 2009,
        month = jul,
       volume = {397},
       number = {1},
        pages = {L51-L54},
          doi = {10.1111/j.1745-3933.2009.00682.x},
archivePrefix = {arXiv},
       eprint = {0905.0637},
 primaryClass = {astro-ph.EP},
       adsurl = {https://ui.adsabs.harvard.edu/abs/2009MNRAS.397L..51K}
}

@ARTICLE{Skopal05,
       author = {{Skopal}, A.},
        title = "{Disentangling the composite continuum of symbiotic binaries. I. S-type systems}",
      journal = {\aap},
         year = 2005,
        month = sep,
       volume = {440},
       number = {3},
        pages = {995-1031},
          doi = {10.1051/0004-6361:20034262},
archivePrefix = {arXiv},
       eprint = {astro-ph/0507272},
 primaryClass = {astro-ph},
       adsurl = {https://ui.adsabs.harvard.edu/abs/2005A&A...440..995S}
}

@ARTICLE{Skopal98,
       author = {{Skopal}, A.},
        title = "{On the nature of apparent changes of the orbital period in symbiotic binaries}",
      journal = {\aap},
         year = 1998,
        month = oct,
       volume = {338},
        pages = {599-611},
       adsurl = {https://ui.adsabs.harvard.edu/abs/1998A&A...338..599S}
}

@ARTICLE{Fekel00_II,
       author = {{Fekel}, Francis C. and {Hinkle}, Kenneth H. and {Joyce}, Richard R. and {Skrutskie}, Michael F.},
        title = "{Infrared Spectroscopy of Symbiotic Stars. II. Orbits for Five S-Type Systems with Two-Year Periods}",
      journal = {\aj},
         year = 2000,
        month = dec,
       volume = {120},
       number = {6},
        pages = {3255-3264},
          doi = {10.1086/316872},
       adsurl = {https://ui.adsabs.harvard.edu/abs/2000AJ....120.3255F}
}

@ARTICLE{Skopal01,
       author = {{Skopal}, A. and {Teodorani}, M. and {Errico}, L. and {Vittone}, A.~A. and {Ikeda}, Y. and {Tamura}, S.},
        title = "{A photometric and spectroscopic study of the eclipsing symbiotic binary AX Persei}",
      journal = {\aap},
         year = 2001,
        month = feb,
       volume = {367},
        pages = {199-210},
          doi = {10.1051/0004-6361:20000413},
       adsurl = {https://ui.adsabs.harvard.edu/abs/2001A&A...367..199S}
}

@ARTICLE{Skopal03,
       author = {{Skopal}, A.},
        title = "{Discovery of the eclipse in the symbiotic binary Z Andromedae}",
      journal = {\aap},
         year = 2003,
        month = apr,
       volume = {401},
        pages = {L17-L20},
          doi = {10.1051/0004-6361:20030332},
archivePrefix = {arXiv},
       eprint = {astro-ph/0304046},
 primaryClass = {astro-ph},
       adsurl = {https://ui.adsabs.harvard.edu/abs/2003A&A...401L..17S}
}

@ARTICLE{Fekel00_I,
       author = {{Fekel}, Francis C. and {Joyce}, Richard R. and {Hinkle}, Kenneth H. and {Skrutskie}, Michael F.},
        title = "{Infrared Spectroscopy of Symbiotic Stars. I. Orbits for Well-Known S-Type Systems}",
      journal = {\aj},
         year = 2000,
        month = mar,
       volume = {119},
       number = {3},
        pages = {1375-1388},
          doi = {10.1086/301260},
       adsurl = {https://ui.adsabs.harvard.edu/abs/2000AJ....119.1375F}
}

@ARTICLE{Luna07,
       author = {{Luna}, G.~J.~M. and {Sokoloski}, J.~L.},
        title = "{The Nature of the Hard X-Ray-Emitting Symbiotic Star RT Cru}",
      journal = {\apj},
         year = 2007,
        month = dec,
       volume = {671},
       number = {1},
        pages = {741-747},
          doi = {10.1086/522576},
archivePrefix = {arXiv},
       eprint = {0708.2576},
 primaryClass = {astro-ph},
       adsurl = {https://ui.adsabs.harvard.edu/abs/2007ApJ...671..741L}
}

@ARTICLE{Collins17,
       author = {{Collins}, Karen A. and {Kielkopf}, John F. and {Stassun}, Keivan G. and {Hessman}, Frederic V.},
        title = "{AstroImageJ: Image Processing and Photometric Extraction for Ultra-precise Astronomical Light Curves}",
      journal = {\aj},
         year = 2017,
        month = feb,
       volume = {153},
       number = {2},
          eid = {77},
        pages = {77},
          doi = {10.3847/1538-3881/153/2/77},
archivePrefix = {arXiv},
       eprint = {1701.04817},
 primaryClass = {astro-ph.IM},
       adsurl = {https://ui.adsabs.harvard.edu/abs/2017AJ....153...77C}
}

@ARTICLE{Lomb76,
       author = {{Lomb}, N.~R.},
        title = "{Least-Squares Frequency Analysis of Unequally Spaced Data}",
      journal = {\apss},
         year = 1976,
        month = feb,
       volume = {39},
       number = {2},
        pages = {447-462},
          doi = {10.1007/BF00648343},
       adsurl = {https://ui.adsabs.harvard.edu/abs/1976Ap&SS..39..447L}
}

@ARTICLE{Scargle82,
       author = {{Scargle}, J.~D.},
        title = "{Studies in astronomical time series analysis. II. Statistical aspects of spectral analysis of unevenly spaced data.}",
      journal = {\apj},
         year = 1982,
        month = dec,
       volume = {263},
        pages = {835-853},
          doi = {10.1086/160554},
       adsurl = {https://ui.adsabs.harvard.edu/abs/1982ApJ...263..835S}
}

@ARTICLE{Vanderplas18,
       author = {{VanderPlas}, Jacob T.},
        title = "{Understanding the Lomb-Scargle Periodogram}",
      journal = {\apjs},
         year = 2018,
        month = may,
       volume = {236},
       number = {1},
          eid = {16},
        pages = {16},
          doi = {10.3847/1538-4365/aab766},
archivePrefix = {arXiv},
       eprint = {1703.09824},
 primaryClass = {astro-ph.IM},
       adsurl = {https://ui.adsabs.harvard.edu/abs/2018ApJS..236...16V}
}

@ARTICLE{Shappee14,
       author = {{Shappee}, B.~J. and {Prieto}, J.~L. and {Grupe}, D. and {Kochanek}, C.~S. and {Stanek}, K.~Z. and {De Rosa}, G. and {Mathur}, S. and {Zu}, Y. and {Peterson}, B.~M. and {Pogge}, R.~W. and {Komossa}, S. and {Im}, M. and {Jencson}, J. and {Holoien}, T.~W. -S. and {Basu}, U. and {Beacom}, J.~F. and {Szczygie{\l}}, D.~M. and {Brimacombe}, J. and {Adams}, S. and {Campillay}, A. and {Choi}, C. and {Contreras}, C. and {Dietrich}, M. and {Dubberley}, M. and {Elphick}, M. and {Foale}, S. and {Giustini}, M. and {Gonzalez}, C. and {Hawkins}, E. and {Howell}, D.~A. and {Hsiao}, E.~Y. and {Koss}, M. and {Leighly}, K.~M. and {Morrell}, N. and {Mudd}, D. and {Mullins}, D. and {Nugent}, J.~M. and {Parrent}, J. and {Phillips}, M.~M. and {Pojmanski}, G. and {Rosing}, W. and {Ross}, R. and {Sand}, D. and {Terndrup}, D.~M. and {Valenti}, S. and {Walker}, Z. and {Yoon}, Y.},
        title = "{The Man behind the Curtain: X-Rays Drive the UV through NIR Variability in the 2013 Active Galactic Nucleus Outburst in NGC 2617}",
      journal = {\apj},
         year = 2014,
        month = jun,
       volume = {788},
       number = {1},
          eid = {48},
        pages = {48},
          doi = {10.1088/0004-637X/788/1/48},
archivePrefix = {arXiv},
       eprint = {1310.2241},
 primaryClass = {astro-ph.HE},
       adsurl = {https://ui.adsabs.harvard.edu/abs/2014ApJ...788...48S}
}

@ARTICLE{Kochanek17,
       author = {{Kochanek}, C.~S. and {Shappee}, B.~J. and {Stanek}, K.~Z. and {Holoien}, T.~W. -S. and {Thompson}, Todd A. and {Prieto}, J.~L. and {Dong}, Subo and {Shields}, J.~V. and {Will}, D. and {Britt}, C. and {Perzanowski}, D. and {Pojma{\'n}ski}, G.},
        title = "{The All-Sky Automated Survey for Supernovae (ASAS-SN) Light Curve Server v1.0}",
      journal = {\pasp},
         year = 2017,
        month = oct,
       volume = {129},
       number = {980},
        pages = {104502},
          doi = {10.1088/1538-3873/aa80d9},
archivePrefix = {arXiv},
       eprint = {1706.07060},
 primaryClass = {astro-ph.SR},
       adsurl = {https://ui.adsabs.harvard.edu/abs/2017PASP..129j4502K}
}

@INPROCEEDINGS{Ricker14,
       author = {{Ricker}, George R. and {Winn}, Joshua N. and {Vanderspek}, Roland and {Latham}, David W. and {Bakos}, G{\'a}sp{\'a}r. {\'A}. and {Bean}, Jacob L. and {Berta-Thompson}, Zachory K. and {Brown}, Timothy M. and {Buchhave}, Lars and {Butler}, Nathaniel R. and {Butler}, R. Paul and {Chaplin}, William J. and {Charbonneau}, David and {Christensen-Dalsgaard}, J{\o}rgen and {Clampin}, Mark and {Deming}, Drake and {Doty}, John and {De Lee}, Nathan and {Dressing}, Courtney and {Dunham}, E.~W. and {Endl}, Michael and {Fressin}, Francois and {Ge}, Jian and {Henning}, Thomas and {Holman}, Matthew J. and {Howard}, Andrew W. and {Ida}, Shigeru and {Jenkins}, Jon and {Jernigan}, Garrett and {Johnson}, John A. and {Kaltenegger}, Lisa and {Kawai}, Nobuyuki and {Kjeldsen}, Hans and {Laughlin}, Gregory and {Levine}, Alan M. and {Lin}, Douglas and {Lissauer}, Jack J. and {MacQueen}, Phillip and {Marcy}, Geoffrey and {McCullough}, P.~R. and {Morton}, Timothy D. and {Narita}, Norio and {Paegert}, Martin and {Palle}, Enric and {Pepe}, Francesco and {Pepper}, Joshua and {Quirrenbach}, Andreas and {Rinehart}, S.~A. and {Sasselov}, Dimitar and {Sato}, Bun'ei and {Seager}, Sara and {Sozzetti}, Alessandro and {Stassun}, Keivan G. and {Sullivan}, Peter and {Szentgyorgyi}, Andrew and {Torres}, Guillermo and {Udry}, Stephane and {Villasenor}, Joel},
        title = "{Transiting Exoplanet Survey Satellite (TESS)}",
    booktitle = {Space Telescopes and Instrumentation 2014: Optical, Infrared, and Millimeter Wave},
         year = 2014,
       editor = {{Oschmann}, Jacobus M., Jr. and {Clampin}, Mark and {Fazio}, Giovanni G. and {MacEwen}, Howard A.},
       series = {Society of Photo-Optical Instrumentation Engineers (SPIE) Conference Series},
       volume = {9143},
        month = aug,
          eid = {914320},
        pages = {914320},
          doi = {10.1117/12.2063489},
archivePrefix = {arXiv},
       eprint = {1406.0151},
 primaryClass = {astro-ph.EP},
       adsurl = {https://ui.adsabs.harvard.edu/abs/2014SPIE.9143E..20R}
}

@ARTICLE{GaiaCollab21,
       author = {{Gaia Collaboration} and {Brown}, A.~G.~A. and {Vallenari}, A. and {Prusti}, T. and {de Bruijne}, J.~H.~J. and {Babusiaux}, C. and {Biermann}, M. and {Creevey}, O.~L. and {Evans}, D.~W. and {Eyer}, L. and et al.},
        title = "{Gaia Early Data Release 3. Summary of the contents and survey properties}",
      journal = {\aap},
         year = 2021,
        month = may,
       volume = {649},
          eid = {A1},
        pages = {A1},
          doi = {10.1051/0004-6361/202039657},
archivePrefix = {arXiv},
       eprint = {2012.01533},
 primaryClass = {astro-ph.GA},
       adsurl = {https://ui.adsabs.harvard.edu/abs/2021A&A...649A...1G}
}

@ARTICLE{Bailer-jones21,
       author = {{Bailer-Jones}, C.~A.~L. and {Rybizki}, J. and {Fouesneau}, M. and {Demleitner}, M. and {Andrae}, R.},
        title = "{VizieR Online Data Catalog: Distances to 1.47 billion stars in Gaia EDR3 (Bailer-Jones+, 2021)}",
      journal = {VizieR Online Data Catalog},
         year = 2021,
        month = feb,
          eid = {I/352},
        pages = {I/352},
       adsurl = {https://ui.adsabs.harvard.edu/abs/2021yCat.1352....0B}
}

@ARTICLE{Horne86,
       author = {{Horne}, J.~H. and {Baliunas}, S.~L.},
        title = "{A Prescription for Period Analysis of Unevenly Sampled Time Series}",
      journal = {\apj},
         year = 1986,
        month = mar,
       volume = {302},
        pages = {757},
          doi = {10.1086/164037},
       adsurl = {https://ui.adsabs.harvard.edu/abs/1986ApJ...302..757H}
}

@ARTICLE{Zechmeister09,
       author = {{Zechmeister}, M. and {K{\"u}rster}, M.},
        title = "{The generalised Lomb-Scargle periodogram. A new formalism for the floating-mean and Keplerian periodograms}",
      journal = {\aap},
         year = 2009,
        month = mar,
       volume = {496},
       number = {2},
        pages = {577-584},
          doi = {10.1051/0004-6361:200811296},
archivePrefix = {arXiv},
       eprint = {0901.2573},
 primaryClass = {astro-ph.IM},
       adsurl = {https://ui.adsabs.harvard.edu/abs/2009A&A...496..577Z}
}

@ARTICLE{Baluev08,
       author = {{Baluev}, R.~V.},
        title = "{Assessing the statistical significance of periodogram peaks}",
      journal = {\mnras},
         year = 2008,
        month = apr,
       volume = {385},
       number = {3},
        pages = {1279-1285},
          doi = {10.1111/j.1365-2966.2008.12689.x},
archivePrefix = {arXiv},
       eprint = {0711.0330},
 primaryClass = {astro-ph},
       adsurl = {https://ui.adsabs.harvard.edu/abs/2008MNRAS.385.1279B}
}

@INPROCEEDINGS{Mikolajewska02,
       author = {{Mikolajewska}, J. and {Kolotilov}, E.~A. and {Shenavrin}, V.~I. and {Yudin}, B.~F.},
        title = "{What powers the nova-like eruptions of symbiotic binaries}",
    booktitle = {The Physics of Cataclysmic Variables and Related Objects},
         year = 2002,
       editor = {{G{\"a}nsicke}, B.~T. and {Beuermann}, K. and {Reinsch}, K.},
       series = {Astronomical Society of the Pacific Conference Series},
       volume = {261},
        month = jan,
        pages = {645},
       adsurl = {https://ui.adsabs.harvard.edu/abs/2002ASPC..261..645M}
}

@ARTICLE{Bollimpalli18,
       author = {{Bollimpalli}, D.~A. and {Hameury}, J.-M. and {Lasota}, J.-P.},
        title = "{Disc instabilities and nova eruptions in symbiotic systems: RS Ophiuchi and Z Andromedae}",
      journal = {\mnras},
         year = 2018,
        month = dec,
       volume = {481},
       number = {4},
        pages = {5422-5435},
          doi = {10.1093/mnras/sty2555},
archivePrefix = {arXiv},
       eprint = {1804.07916},
 primaryClass = {astro-ph.HE},
       adsurl = {https://ui.adsabs.harvard.edu/abs/2018MNRAS.481.5422B}
}

@ARTICLE{Mikolajewska95,
       author = {{Mikolajewska}, Joanna and {Kenyon}, Scott J. and {Mikolajewski}, Maciej and {Garcia}, Michael R. and {Polidan}, Ronald S.},
        title = "{Evolution of the Symbiotic Binary System AG Draconis}",
      journal = {\aj},
         year = 1995,
        month = mar,
       volume = {109},
        pages = {1289},
          doi = {10.1086/117361},
       adsurl = {https://ui.adsabs.harvard.edu/abs/1995AJ....109.1289M}
}

@ARTICLE{Munari19,
       author = {{Munari}, Ulisse},
        title = "{The Symbiotic Stars}",
      journal = {arXiv e-prints},
         year = 2019,
        month = sep,
          eid = {arXiv:1909.01389},
        pages = {arXiv:1909.01389},
          doi = {10.48550/arXiv.1909.01389},
archivePrefix = {arXiv},
       eprint = {1909.01389},
 primaryClass = {astro-ph.SR},
       adsurl = {https://ui.adsabs.harvard.edu/abs/2019arXiv190901389M}
}

@ARTICLE{Kato23,
       author = {{Kato}, Mariko and {Hachisu}, Izumi},
        title = "{Theoretical Light-curve Models of the Symbiotic Nova CN Cha-Optical Flat Peak for 3 Yr}",
      journal = {\apj},
         year = 2023,
        month = jul,
       volume = {951},
       number = {2},
          eid = {128},
        pages = {128},
          doi = {10.3847/1538-4357/acdb4c},
archivePrefix = {arXiv},
       eprint = {2306.01288},
 primaryClass = {astro-ph.SR},
       adsurl = {https://ui.adsabs.harvard.edu/abs/2023ApJ...951..128K}
}

@ARTICLE{Merc19,
       author = {{Merc}, J. and {G{\'a}lis}, R. and {Wolf}, M. and {Leedj{\"a}rv}, L. and {Teyssier}, F.},
        title = "{The activity of the symbiotic binary Z Andromedae and its latest outburst}",
      journal = {Open European Journal on Variable Stars},
         year = 2019,
        month = apr,
       volume = {197},
        pages = {23},
          doi = {10.48550/arXiv.1905.04251},
archivePrefix = {arXiv},
       eprint = {1905.04251},
 primaryClass = {astro-ph.SR},
       adsurl = {https://ui.adsabs.harvard.edu/abs/2019OEJV..197...23M}
}

@ARTICLE{Merc24,
       author = {{Merc}, J. and {Beck}, P.~G. and {Mathur}, S. and {Garc{\'\i}a}, R.~A.},
        title = "{Accretion-induced flickering variability among symbiotic stars from space photometry with NASA TESS}",
      journal = {\aap},
         year = 2024,
        month = mar,
       volume = {683},
          eid = {A84},
        pages = {A84},
          doi = {10.1051/0004-6361/202348116},
archivePrefix = {arXiv},
       eprint = {2312.16126},
 primaryClass = {astro-ph.SR},
       adsurl = {https://ui.adsabs.harvard.edu/abs/2024A&A...683A..84M}
}

@ARTICLE{Merc25_rew,
       author = {{Merc}, Jaroslav},
        title = "{Symbiotic Stars in the Era of Modern Ground- and Space-Based Surveys}",
      journal = {Galaxies},
         year = 2025,
        month = apr,
       volume = {13},
       number = {3},
          eid = {49},
        pages = {49},
          doi = {10.3390/galaxies13030049},
archivePrefix = {arXiv},
       eprint = {2504.16825},
 primaryClass = {astro-ph.SR},
       adsurl = {https://ui.adsabs.harvard.edu/abs/2025Galax..13...49M}
}

@ARTICLE{Merc25,
       author = {{Merc}, Jaroslav and {Boffin}, Henri M.~J.},
        title = "{Revisiting symbiotic binaries with interferometry: I. The PIONIER archival collection}",
      journal = {\aap},
         year = 2025,
        month = mar,
       volume = {695},
          eid = {A61},
        pages = {A61},
          doi = {10.1051/0004-6361/202553789},
archivePrefix = {arXiv},
       eprint = {2502.04089},
 primaryClass = {astro-ph.SR},
       adsurl = {https://ui.adsabs.harvard.edu/abs/2025A&A...695A..61M}
}

@software{lightkurve18,
       author = {{Lightkurve Collaboration} and {Cardoso}, Jos{\'e} Vin{\'\i}cius de Miranda and {Hedges}, Christina and {Gully-Santiago}, Michael and {Saunders}, Nicholas and {Cody}, Ann Marie and {Barclay}, Thomas and {Hall}, Oliver and {Sagear}, Sheila and {Turtelboom}, Emma and {Zhang}, Johnny and {Tzanidakis}, Andy and {Mighell}, Ken and {Coughlin}, Jeff and {Bell}, Keaton and {Berta-Thompson}, Zach and {Williams}, Peter and {Dotson}, Jessie and {Barentsen}, Geert},
        title = "{Lightkurve: Kepler and TESS time series analysis in Python}",
 howpublished = {Astrophysics Source Code Library, record ascl:1812.013},
         year = 2018,
        month = dec,
          eid = {ascl:1812.013},
archivePrefix = {ascl},
       eprint = {1812.013},
       adsurl = {https://ui.adsabs.harvard.edu/abs/2018ascl.soft12013L}
}

@software{Brasseur19,
       author = {{Brasseur}, C.~E. and {Phillip}, Carlita and {Fleming}, Scott W. and {Mullally}, S.~E. and {White}, Richard L.},
        title = "{Astrocut: Tools for creating cutouts of TESS images}",
 howpublished = {Astrophysics Source Code Library, record ascl:1905.007},
         year = 2019,
        month = may,
          eid = {ascl:1905.007},
archivePrefix = {ascl},
       eprint = {1905.007},
       adsurl = {https://ui.adsabs.harvard.edu/abs/2019ascl.soft05007B}
}

@ARTICLE{magdolen23,
       author = {{Magdolen}, J. and {Dobrotka}, A. and {Orio}, M. and {Miko{\l}ajewska}, J. and {Vanderburg}, A. and {Monard}, B. and {Aloisi}, R. and {Bez{\'a}k}, P.},
        title = "{Recurrent mini-outbursts and a magnetic white dwarf in the symbiotic system FN Sgr}",
      journal = {\aap},
         year = 2023,
        month = jul,
       volume = {675},
          eid = {A140},
        pages = {A140},
          doi = {10.1051/0004-6361/202345935},
archivePrefix = {arXiv},
       eprint = {2306.05095},
 primaryClass = {astro-ph.SR},
       adsurl = {https://ui.adsabs.harvard.edu/abs/2023A&A...675A.140M}
}

@ARTICLE{formiggini09,
       author = {{Formiggini}, Liliana and {Leibowitz}, Elia M.},
        title = "{Discovery of the 1.80 h spin period of the white dwarf of the symbiotic system BF Cyg}",
      journal = {\mnras},
         year = 2009,
        month = jul,
       volume = {396},
       number = {3},
        pages = {1507-1512},
          doi = {10.1111/j.1365-2966.2009.14835.x},
archivePrefix = {arXiv},
       eprint = {0903.5157},
 primaryClass = {astro-ph.SR},
       adsurl = {https://ui.adsabs.harvard.edu/abs/2009MNRAS.396.1507F}
}

@ARTICLE{demartino20,
       author = {{de Martino}, D. and {Bernardini}, F. and {Mukai}, K. and {Falanga}, M. and {Masetti}, N.},
        title = "{Hard X-ray cataclysmic variables}",
      journal = {Advances in Space Research},
         year = 2020,
        month = sep,
       volume = {66},
       number = {5},
        pages = {1209-1225},
          doi = {10.1016/j.asr.2019.09.006},
archivePrefix = {arXiv},
       eprint = {1909.06306},
 primaryClass = {astro-ph.HE},
       adsurl = {https://ui.adsabs.harvard.edu/abs/2020AdSpR..66.1209D}
}

@ARTICLE{Leibowitz08,
       author = {{Leibowitz}, Elia M. and {Formiggini}, Liliana},
        title = "{Activity cycle of the giant star of Z Andromedae and its spin period}",
      journal = {\mnras},
         year = 2008,
        month = mar,
       volume = {385},
       number = {1},
        pages = {445-452},
          doi = {10.1111/j.1365-2966.2008.12847.x},
archivePrefix = {arXiv},
       eprint = {0712.2120},
 primaryClass = {astro-ph},
       adsurl = {https://ui.adsabs.harvard.edu/abs/2008MNRAS.385..445L}
}

@ARTICLE{Kenyon82,
       author = {{Kenyon}, S.~J.},
        title = "{The orbital period of the symbiotic star AX Per.}",
      journal = {\pasp},
         year = 1982,
        month = feb,
       volume = {94},
        pages = {165-168},
          doi = {10.1086/130957},
       adsurl = {https://ui.adsabs.harvard.edu/abs/1982PASP...94..165K}
}

@INPROCEEDINGS{Mikolajewska96,
       author = {{Mikolajewska}, J.},
        title = "{Evolution of an accretion disk in the symbiotic binary CI CYG}",
    booktitle = {IAU Colloquium 158: Cataclysmic Variables and Related Objects},
         year = 1996,
       editor = {{Evans}, A. and {Wood}, Janet H.},
       series = {Astrophysics and Space Science Library},
       volume = {208},
        month = jan,
        pages = {335},
          doi = {10.1007/978-94-009-0325-8_102},
       adsurl = {https://ui.adsabs.harvard.edu/abs/1996ASSL..208..335M}
}

@ARTICLE{Formiggini90,
       author = {{Formiggini}, L. and {Leibowitz}, E.~M.},
        title = "{A reflection model for the light curves of three symbiotic stars.}",
      journal = {\aap},
         year = 1990,
        month = jan,
       volume = {227},
        pages = {121-129},
       adsurl = {https://ui.adsabs.harvard.edu/abs/1990A&A...227..121F}
}
\bibliographystyle{aasjournalv7}



\end{document}